\documentclass[%
 aip,
 amsmath,amssymb,
 reprint,%
]{revtex4-1}

\usepackage{graphicx}
\usepackage{dcolumn}
\usepackage{bm}

\usepackage[utf8]{inputenc}
\usepackage[T1]{fontenc}
\usepackage{mathptmx}
\usepackage{etoolbox}

\usepackage{siunitx}
\usepackage[dvipsnames]{xcolor}
\definecolor{darkblue}{rgb}{0,0,0.6}
\definecolor{darkred}{rgb}{0.6,0,0}
\usepackage[colorlinks=true,urlcolor=darkblue,citecolor=darkblue, linkcolor=darkred,hyperfootnotes=false]{hyperref}

\newcommand{\dd}{\text{d}}

\newcommand{\EE}{\boldsymbol{E}}

\newcommand{\kk}{\boldsymbol{k}}

\newcommand{\MM}{\boldsymbol{M}}

\newcommand{\mcO}{\mathcal{O}}

\newcommand{\rr}{\boldsymbol{r}}

\newcommand{\uu}{\boldsymbol{u}}

\renewcommand{\vec}[1]{\boldsymbol{#1}}

\newcommand{\ind}[1]{_{\mathrm{#1}}}

\DeclareSIUnit{\molar}{M}
\DeclareSIUnit{\mmol}{\milli \molar}
\DeclareSIUnit{\vpnm}{\volt\per\nano\meter}

\newcommand{\Ec}{E_c}
\newcommand{\Ew}{E_w}

\newcommand{\rhod}{\rho_d}

\newcommand{\dse}{\Delta\sigma\ind{el}}
\newcommand{\dsh}{\Delta\sigma\ind{hyd}}

\makeatletter
\def\@email#1#2{%
 \endgroup
 \patchcmd{\titleblock@produce}
  {\frontmatter@RRAPformat}
  {\frontmatter@RRAPformat{\produce@RRAP{*#1\href{mailto:#2}{#2}}}\frontmatter@RRAPformat}
  {}{}
}%
\makeatother
\begin{document}

\preprint{AIP/123-QED}

\title{Nonlinear conductivity of aqueous electrolytes: beyond the first Wien effect}

\author{Hélène Berthoumieux}
\email{helene.berthoumieux@espci.fr}
\affiliation{UMR CNRS Gulliver 7083, ESPCI Paris, PSL Research University, 75005 Paris, France}%

\author{Vincent Démery}%
\email{vincent.demery@espci.psl.eu}
\affiliation{UMR CNRS Gulliver 7083, ESPCI Paris, PSL Research University, 75005 Paris, France}%
\affiliation{Univ Lyon, ENSL, CNRS, Laboratoire de Physique, F-69342 Lyon, France}

\author{Anthony C. Maggs}%
\email{anthony.maggs@espci.fr}
\affiliation{UMR CNRS Gulliver 7083, ESPCI Paris, PSL Research University, 75005 Paris, France}%

%
%
%

\date{\today}

\begin{abstract}
The conductivity of strong electrolytes increases under high electric fields, a nonlinear response known as the first Wien effect. 
Here, using molecular dynamics simulations we show that this increase is almost suppressed in moderately concentrated aqueous electrolytes due to the alignment of the water molecules by the electric field.
As a consequence of this alignement, the permittivity of water decreases and becomes anisotropic, an effect which can be measured in simulations and reproduced by a model of water molecules as dipoles.
We incorporate the resulting anisotropic interactions between the ions into a Stochastic Density Field Theory and calculate  ionic correlations as well as  corrections to the Nernst-Einstein conductivity, which are in qualitative agreement with the numerical simulations.
\end{abstract}

\maketitle

\section{Introduction}

Almost a century ago, Wien probed the response of strong electrolytes under very high electric fields, up to $\qty{0.05}{\volt\per\nano\meter}$. 
He demonstrated a generic increase of the conductivity upon increasing the
electric field, and later found an increase for weak electrolytes; these two
phenomena are the first and second Wien effects~\cite{Eckstrom1939Wien}.
The first Wien effect was explained by Wilson~\cite{Wilson1936}, and later Onsager and Kim, who extended the Debye-Hückel-Onsager (DHO) theory~\cite{Debye1923, Onsager1927} to finite electric fields~\cite{Onsager1957Wien}.
According to the DHO theory,  ions are surrounded by a cloud of counterions, which is distorted under an external electric field.
This distorted cloud generates an additional drag on the ions and reduces their mobility, thereby reducing the conductivity of the solution.
Wilson, Onsager and Kim showed that large electric fields destroy the ionic cloud, thus restoring the bare mobility of the ions, which leads to an increase in the conductivity compared to the small field limit.
The second Wien effect is due to the increase of the dissociation constant of a weak electrolyte under a large electric field and was explained by Onsager~\cite{Onsager1934}.
Today, fields of up to $\qty{0.1}{\volt\per\nano\meter}$ are routinely applied in nanofluidic devices involving atomically thin membranes~\cite{Feng2016Observation, Cai2022Wien}, electrodes~\cite{Montenegro2021} or channels~\cite{Robin2023Long-term}.
Such high fields gave experimental access to the second Wien effect for water dissociation~\cite{Cai2022Wien} and are also used for the investigation of other ionic transport phenomena including the ionic Coulomb blockade~\cite{Feng2016Observation}, or the creation of ion-based memristors in confined geometries~\cite{Robin2021Modeling,Robin2023Long-term}.
These experiments call for a better theoretical understanding of the behavior of electrolytes under large electric fields.

\begin{figure}
  \begin{center}
    \includegraphics[width=.95\linewidth]{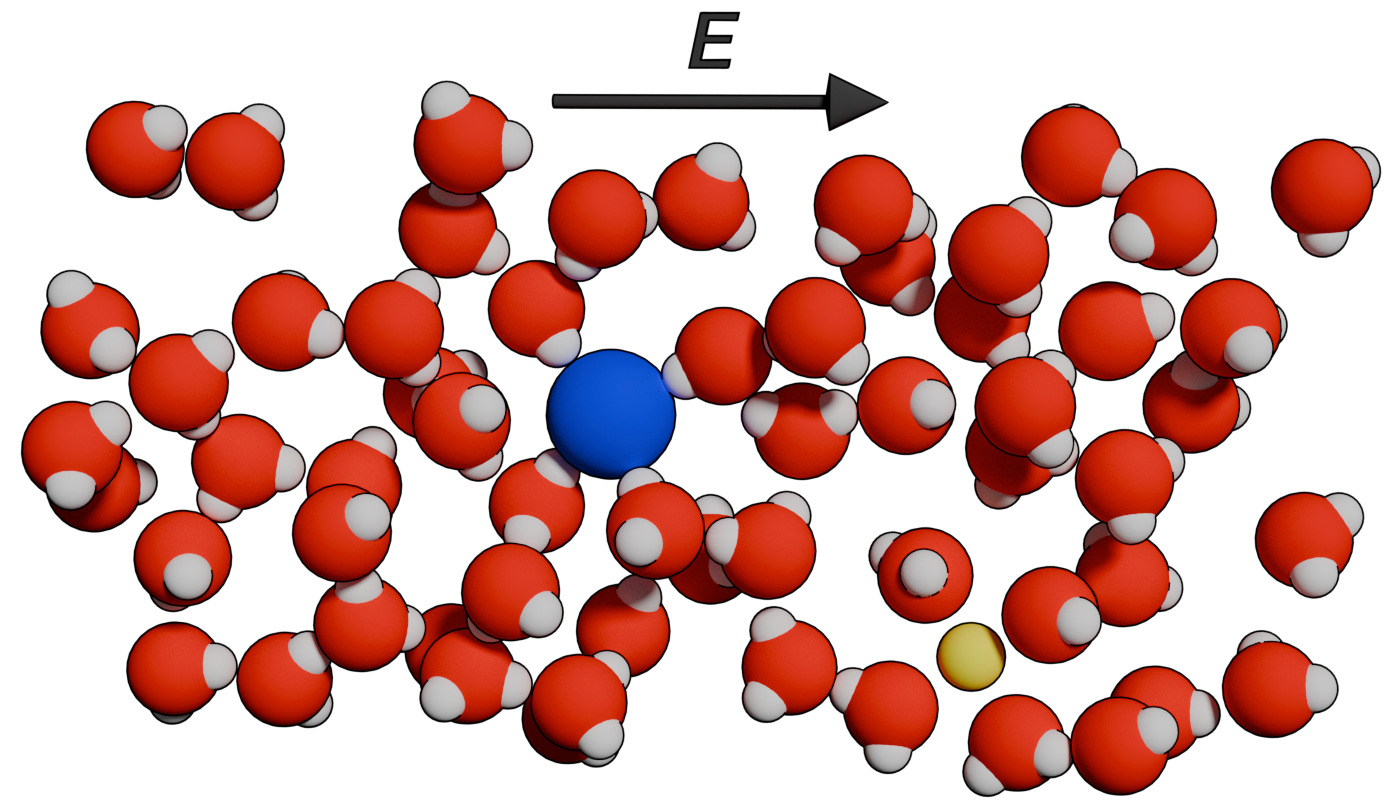}
    \caption{Snapshot of the classical MD simulations with an electric field of $\qty{0.5}{\vpnm}$. The sodium and chloride ions are represented by a yellow and a blue sphere, respectively. The orientation of the water molecules by the external field (black arrow) is clearly visible.}
    \label{fig:schematic}
  \end{center}
\end{figure}

Recently, the response of strong aqueous electrolytes has been investigated using classical molecular dynamics (MD) simulations with an explicit description for water~\cite{Lesnicki2020, Lesnicki2021}. 
Unexpectedly, the conductivity was found to be a constant function of the electrostatic field, up to electric fields of $\qty{1}{\volt\per\nano\meter}$.
Applying the electric field on the ions only, and not on the water molecules, a large increase in the conductivity has been recovered~\cite{Lesnicki2020, Lesnicki2021}.
These results suggest that the molecular properties of water also affect the nonlinear response of electrolytes.
However, most of the theoretical efforts from the DHO theory, such as its recent reformulation using Stochastic Density Field Theory (SDFT) and its extensions~\cite{Dean1996, Demery2016Conductivity, Peraud2017, Donev2019, Avni2022Conductivity, Avni2022Conductance, Bernard2023Analytical} or approaches based on the Mean Spherical Approximation~\cite{Dufreche2005Analytical}, have focused on a better description of the ions, leaving aside the molecular properties of the solvent.
The external field polarizes water by aligning the dipoles of water molecules (Fig.~\ref{fig:schematic}) and affects two important water properties for ion transport, making them anisotropic: its permittivity~\cite{Booth1951Dielectric, Piekara1958, Yeh1999Dielectric, Kotodziej1975High} and its viscosity~\cite{Andrade1946,Lyklema1961,Zong2016Viscosity,Jin2022Direct}.
Classical MD simulations report that even at high field strengths of $\qty{1}{\vpnm}$, the increase in viscosity remains moderate~\cite{Jin2022Direct}, while the permittivity decreases by an order of magnitude~\cite{Yeh1999Dielectric}.

Here we show that the field-induced decrease in water permittivity leads to an increase in ion-ion interactions and compensates for the Debye cloud destructuring. 
In addition, we find that for a certain range of electrostatic fields and electrolyte concentrations, the cloud enhancement effect induced by the permittivity drop is dominant, resulting in a decreasing conductivity.
We use a model of water as dipoles on a lattice~\cite{Booth1951Dielectric, Abrashkin2007Dipolar} to encode the molecular properties of water in a field-dependent, anisotropic, permittivity tensor.
We then incorporate this permittivity tensor into SDFT to obtain the ionic correlations and the conductivity of the solution, which is nonmonotonic in the electric field.
We show that the nonlinear response is controlled by the interplay of two characteristic electric fields: the electric field $\Ec$ where the ionic cloud is destroyed for a constant permittivity, and the electric field $\Ew$ needed to orient the water molecules.
Last, while the main effect is due to the permittivity drop, the permittivity anisotropy significantly affects the conductivity.

\section{Conductivity from Molecular Dynamics simulations}

\begin{figure}
  \begin{center}
    \includegraphics[scale=1]{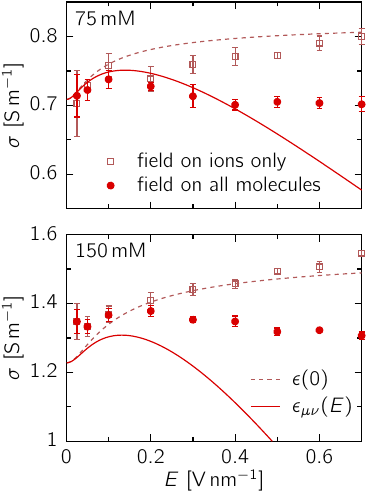}
    \caption{Conductivity as a function of the electric field for aqueous solutions of NaCl at concentrations $\rho=\qtylist{75;150}{\milli\molar}$ from numerical simulations (symbols) and theory (lines). 
    Open symbols correspond to simulations where the external electric field is only applied to the ions, while it is applied to the ions and the water molecules for closed symbols. 
    Dashed lines are the DHO theory, with a constant water relative permittivity of $\epsilon_r=71$; solid lines are the theory described here, which includes the effect of the electric field on the permittivity (Eqs.~(\ref{eq:correc_el}, \ref{eq:correc_hyd})).
    The Nernst-Einstein conductivity is $\sigma_0\simeq\qty{0.94}{\siemens\per\meter}$ for $\rho=\qty{75}{\mmol}$ and $\sigma_0\simeq\qty{1.88}{\siemens\per\meter}$ for $\rho=\qty{150}{\mmol}$.
}
    \label{fig:conductivity}
  \end{center}
\end{figure}

We ran classical Molecular Dynamics simulations to measure the conductivity of aqueous solutions of NaCl, using the rigid, non-polarizable SPC/E model for water~\cite{Berendsen1987Missing, Smith1993Viscosity, Loche2021Transferable}, at moderate concentrations $\rho=\qtylist{75;150}{\milli\molar}$ and at temperature $T=\qty{300}{\kelvin}$~\cite{Hess2008Gromacs4, GitHub} (App.~\ref{app:md}). 
We measured the conductivity for electric fields ranging from $\qtyrange{0.025}{0.7}{\vpnm}$. 
At these field amplitudes, some of the water molecules align their dipole moment along the field, but the system retains its liquid structure and the field does not disrupt the ion solvation layers (Fig.~\ref{fig:schematic}).
The conductivity as a function of $E$ is presented by filled circles in Fig.~\ref{fig:conductivity}. 
At both concentrations, the conductivity dependence on the electric field is subtle: it slightly increases and then decreases with the electric field, with a maximum around $\qty{0.1}{\vpnm}$.
To probe the effect of the alignment of the water molecules on the external field, we performed simulations where the external field has only been applied to the ions~\cite{Lesnicki2021} (Fig.~\ref{fig:conductivity}, open squares).
In this configuration, the conductivity increases over the whole range of applied electric fields, corresponding to the first Wien effect.
For both concentrations, the data are quantitatively reproduced by the DHO theory, without any fitting parameter (Fig.~\ref{fig:conductivity}, dashed lines).

\section{Anisotropic field-dependent permittivity}

From the observations above, we conclude that the nonlinear conductivity is strongly affected by the response of the water molecules to the external field.
Indeed, the external field polarizes water (Fig.~\ref{fig:schematic}), thereby reducing the response of water molecules to extra fields such as the ones generated by the ions, that is, reducing the permittivity of the medium. 
Computing this effect from first principles has proven a difficult task~\cite{Fulton2009Nonlinear}.
Instead, we use a model where the water molecules are represented by dipoles on a lattice with magnitude $p$ and density $\rhod$~\cite{Booth1951Dielectric, Abrashkin2007Dipolar}, and allow ourselves to fit these quantities~\cite{monetJCP2021}.
Note that the role of the lattice is to enforce the incompressibility of water.
Neglecting the interactions between the dipoles, the free energy density under an external field with magnitude $E$ is
\begin{equation}\label{eq:free_energy_density}
f(E) = -\frac{\epsilon_0}{2}E^2-\frac{\rhod}{\beta}\log \left(\frac{\sinh(\beta p E)}{\beta p E} \right),
\end{equation}
where $\epsilon_0$ is the vacuum permittivity and $\beta$ is the inverse thermal energy.
While Ref.~\onlinecite{Abrashkin2007Dipolar} focused on one-dimensional systems, the tensorial permittivity may be obtained as $\epsilon_{\mu\nu}=-\partial^2 f(E)/(\partial E_\mu\partial E_\nu)$~\cite{landau2013electrodynamics}.
Here, the relevant permittivity to describe the interaction between the ions is given by the expansion of this relation around the external field, leading to a diagonal permittivity tensor with different values along the external field and transverse to it (App.~\ref{app:aniso}):
\begin{align}
\epsilon_\parallel &  = \epsilon_0+\rhod\beta p^2 \left[\frac{1}{x^2}-\frac{1}{\sinh(x)^2} \right], \label{eq:eps_para}\\
\epsilon_\perp & = \epsilon_0+\rhod\beta p^2\left[\frac{\coth(x)}{x}-\frac{1}{x^2} \right],\label{eq:eps_perp}
\end{align}
where $x=\beta p E=E/\Ew$ is the dimensionless field, $\Ew=1/(\beta p)$ being the external field necessary to orient a water molecule.
Both expressions give $\epsilon_0+\beta \rhod p^2/3=\epsilon_0\epsilon_r$  for small fields and decay to $\epsilon_0$ for large fields.

\begin{figure}
  \begin{center}
    \includegraphics[scale=1]{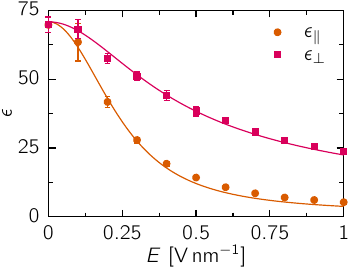}
    \caption{Longitudinal and transverse permittivity from numerical simulations (App.~\ref{app:md}, symbols) and theory (solid lines, Eqs.~(\ref{eq:eps_para},\ref{eq:eps_perp})).}
    \label{fig:permittivity}
  \end{center}
\end{figure}

We simulated a solution of pure water under an increasing electrostatic field from $0$ to $\qty{1}{\vpnm}$ applied along the $x$ direction. 
We computed the permittivity tensor from the fluctuations of the total dipole moment $\vec{M}$ of the simulation cell (App.~\ref{app:md_perm}). 
The longitudinal permittivity $\epsilon_\parallel$ is derived from the fluctuations of $M_x$ and the transverse permittivity $\epsilon_\perp$ from the fluctuations of $M_y$ and $M_z$. 
	Analytical expressions fit very well to the longitudinal and transverse permittivities measured in the simulations with $p=\qty{3.6e-29}{\coulomb\meter}$ and $\rhod=\qty{6.0}{\nano\meter^{-3}}$ (Fig.~\ref{fig:permittivity}), that is, with a dipole moment that is about 5 times larger than that of the water molecule, consistent with the calculation of Booth~\cite{Booth1951Dielectric}, and a density that is about 5 times smaller than that of the liquid.
With this value of the dipolar moment $p$, the field necessary to orient a water molecule is $\Ew\simeq\qty{0.12}{\vpnm}$.
Beyond the decay, the permittivity tensor exhibits a strong anisotropy at intermediary fields, the transverse component being much larger than the longitudinal one.

\section{Conductivity from Stochastic Density Field Theory}

The salt reduces the permittivity of water~\cite{hasted48,levy2012,seal2019}, but at the moderate concentrations we consider the decrease is limited to a few percent. 
We therefore assume that the permittivity in electrolytes can be approximated to that of pure water (Fig.~\ref{fig:permittivity}). 
Its anisotropy results in anisotropic electrostatic interactions between the ions: the pairwise interaction between ionic species $i$ and $j$ is given by $\tilde V_{ij}(\kk)=q^2 z_i z_j/(\epsilon_\parallel k_\parallel^2+\epsilon_\perp k_\perp^2)$ in Fourier space, where $q$ is the elementary charge and $z_i$ is the valency of the ions of type $i$, $\kk$ is the wavenumber and $k_\parallel$ and $\kk_\perp$ are its components parallel and perpendicular to the external field, respectively.
We resort to SDFT to compute the ionic correlations.
The ions are represented as point particles obeying an overdamped Langevin
dynamics with mobilities $\kappa_i$ under the action of the interparticle forces calculated from $V_{ij}$ and the external field $\EE$.
Using SDFT, the microscopic dynamics of the ions is mapped onto an overdamped Langevin dynamics for the density field of each species~\cite{Dean1996}.
Assuming small density fluctuations, which corresponds to the Debye-Hückel approximation, the density dynamics can be linearized and the correlations can be computed~\cite{Demery2016Conductivity} (App.~\ref{app:correl}).

The Nernst-Einstein conductivity $\sigma_0=\rho q^2\kappa$, where $\kappa = \kappa_++\kappa_-$, is reduced by the electrostatic $\dse$ and hydrodynamic $\dsh$ corrections, which are deduced from the ionic correlations~\cite{Demery2016Conductivity, Avni2022Conductivity, Bonneau2023Temporal}:
\begin{align}
  \frac{\Delta\sigma_\mathrm{el}}{\sigma_0}&
  = -\frac{\beta q^2 m }{4\sqrt{2}\pi\epsilon_0\epsilon_r}
\int_0^1 \dd y \frac{y^2 \gamma_y}{\alpha_y^{3/2}}, \label{eq:correc_el}\\
\frac{\Delta\sigma\ind{hyd}}{\sigma_0}&
=-\frac{m}{2\sqrt{2}\pi \eta\kappa}\int_0^1\dd y\frac{(1-y^2)(1+\gamma_y)}{\sqrt{\alpha_y}}. \label{eq:correc_hyd}
\end{align}
We have introduced the inverse Debye length at zero field $m = \sqrt{2\beta q^2\rho/(\epsilon_0\epsilon_r)}$. 
The anisotropic permittivity is encoded in $\alpha_y = [\epsilon_\parallel y^2+\epsilon_\perp (1-y^2)]/(\epsilon_0\epsilon_r)$ and $\gamma_y=\left[1+\sqrt{2(1+F^2y^2\alpha_y)}\right]^{-1}$.
The dimensionless electric field is defined by $F=E/\Ec$, where $\Ec=m/(\beta q)$ is the field necessary to destroy the ionic cloud when the permittivity is assumed to be constant.
The ionic mobilities $\kappa_i$ are measured via the Einstein relation in simulations of very dilute solutions without external field (App.~\ref{app:md_ionmob}).
The hydrodynamic correction involves the viscosity $\eta$ of the solvent, which we assume to be constant and isotropic; we use the value $\eta=\qty{0.729e-3}{\pascal\second}$~\cite{Smith1993Viscosity}. 
These expressions hold for monovalent salts; the general results are given in App.~\ref{app:elec} and \ref{app:hydro}.

Our prediction for the conductivity, $\sigma_0+\dse+\dsh$, is shown by red solid lines in Fig.~\ref{fig:conductivity}.
For comparison, the DHO prediction, which assumes a constant isotropic permittivity $\epsilon_0\epsilon_r$, is shown as light dashed lines.
Up to electric fields of  $\qty{0.4}{\vpnm}$, the agreement is quantitative for the smaller concentration, $\rho=\qty{75}{\milli\molar}$; 
for the larger concentration, $\rho=\qty{150}{\milli\molar}$, the theory underestimates the conductivity, as expected~\cite{Avni2022Conductivity}, but the agreement is still qualitative.
The conductivity predicted by DHO is an increasing function of the electric field, as measured in the simulations where the external field acts only on the ions.
In contrast, our calculation predicts a nonmonotonic dependence, which is compatible with the simulations where the external field acts on all the molecules; in particular, the electric field where the conductivity is predicted to be maximal corresponds to the location of the slight maximum observed in simulations.
For large electric fields, the discrepancy between theory and simulations increases: as the permittivity decays, the correlations between the ions increase and the assumption of small density fluctuations, used to compute the correlations no longer holds.

\begin{figure}
  \begin{center}
    \includegraphics[scale=1]{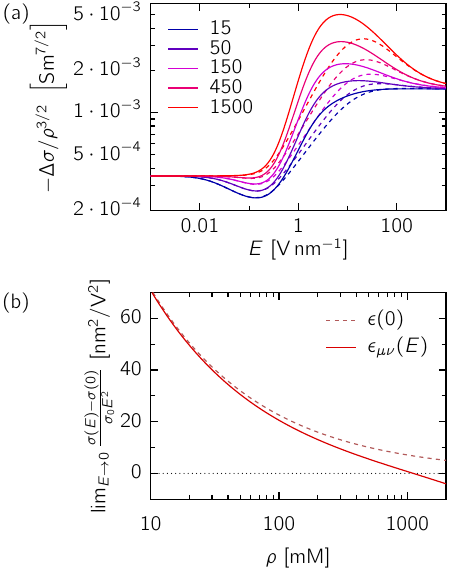}
    \caption{(a) Rescaled correction to the Nernst-Einstein conductivity for different concentrations (indicated in $\unit{\mmol}$), with the anisotropic permittivity (solid lines) and the isotropic permittivity $\epsilon=\epsilon_{\mu\mu}/3$ (dashed lines).
    (b) Coefficient of $E^2$ in the small field expansion as a function of the salt concentration.}
    \label{fig:correction}
  \end{center}
\end{figure}

To better understand the shape of the conductivity curve, we plot the theoretical prediction for the total correction, $\Delta\sigma=\dse+\dsh$, as solid lines for a wide range of concentrations and a wide range of external fields, up to unrealistic values, in Fig.~\ref{fig:correction}(a).
For the most dilute system, $\rho=\qty{15}{\milli\molar}$, the correction first decays, which corresponds to an increase of the conductivity, and then increases; for the densest system, $\rho=\qty{1.5}{\molar}$, the contrary is observed: the correction first increases and then decays.
In the intermediate range, for instance for $\rho=\qty{150}{\milli\molar}$, the correction decreases, increases and decreases again; the last decay is not observed in Fig.~\ref{fig:conductivity} as it occurs for large values of the external field, $E\gtrsim\qty{10}{\vpnm}$.
Two nonlinear effects compete here; one is the decay of the water permittivity, which enhances the interactions and the ionic correlations, occurring at the characteristic field $\Ew$.
The other is the destruction of the ionic correlations above the characteristic field $\Ec$.
For weak electric fields, the evolution of the conductivity is the sum of the two effects (App.~\ref{app:elec_small} and \ref{app:hydro_small}):
\begin{equation}\label{eq:weak_field_expansion}
  \frac{\sigma(E)-\sigma(0)}{\sigma_0} \underset{E\to 0}{\sim} a \frac{E^2}{\Ec^2} + b \frac{E^2}{\Ew^2}.
\end{equation}
The coefficients $a$ and $b$ scale as $\sqrt{\rho}$ and have opposite sign, $a>0$ while $b<0$.
The characteristic field $\Ec$ also scales as $\sqrt{\rho}$, while $\Ew$ is a characteristic of pure water and does not depend on the salt concentration. 
As a consequence, the coefficient of $E^2$ in Eq.~(\ref{eq:weak_field_expansion}) may change sign as a function of the salt concentration.
This is shown in Figure ~\ref{fig:correction}(b) where we plot this coefficient as a function of $\rho$ for solutions associated with field-dependent permittivity (solid line) versus solutions with constant permittivity (dashed line). The change of sign occurs for concentrated solutions ($\rho \geq \qty{1}{\molar}$), however note that the correction due to the decrease of the permittivity is barely visible for concentrations below $\qty{100}{\mmol}$.

The nonmonotonic evolution of the conductivity with the external field is chiefly explained by the interplay of the destruction of the ionic correlations and the decay of the permittivity, irrespective of its anisotropy.
However, considering an isotropic medium by replacing the anisotropic permittivity $\epsilon_{\mu\nu}$ by its angular averaged value $\epsilon_{\mu\mu}/3$ significantly reduces the correction $\Delta\sigma$, which is shown by the dashed lines in Fig.~\ref{fig:correction}.

\section{Conclusion}

We have shown that the conductivity of aqueous electrolytes may be a nonmonotonic function of the external field, and we have explained the result by the evolution of the permittivity of water, which decreases with field strength and also becomes anisotropic.
The decrease as well as  the anisotropy of the permittivity can be included in SDFT to reproduce qualitatively the evolution of the conductivity, and even yield quantitative predictions at low concentrations.

In the future, this integration of molecular properties of water into the analytically tractable framework of SDFT could be extended to include, for instance, a non-local permittivity $\tilde \epsilon(\kk)$ to account for the finite size of the water molecules~\cite{Bopp1996,Berthoumieux2018Gaussian}, or a frequency-dependent permittivity $\tilde \epsilon(\omega)$ to account for their relaxational dynamics~\cite{Bopp1998,Illien2024Stochastic}.
Finally, beyond the bulk conductivity of strong electrolytes, we expect that the molecular properties of water also affect ionic transport in quasi two- or one-dimensional systems, where confinement and ions control the structure of water~\cite{Jalali2021}.
This results in a stronly anisotropic permittivity tensor, which values can be significantly higher~\cite{Loche2019Giant,Wang2024} or lower~\cite{fumagalli2018,Jalali2021} than the bulk one.
The inclusion of the molecular properties of the solvent in SDFT would thus allow a better description of the first and second Wien effects in such systems~\cite{Robin2021Modeling, Robin2023Long-term}.

\appendix

\section*{Data Availability Statement}

The data that support the findings of this study are available from the corresponding author upon reasonable request.
The simulation files and the codes used to interpret the simulations results are openly available in \href{https://github.com/acmaggs/Wien}{GitHub} (Ref.~\onlinecite{GitHub}).

\section{Molecular dynamics simulations}
\label{app:md}

\subsection{Simulation methods}
\label{app:md_methods}

Simulations are performed using the GROMACS 2021 molecular dynamics simulation package~\cite{Hess2008Gromacs4}.
Simulation boxes are periodically replicated in all directions, and long-range electrostatics are handled using the smooth particle mesh Ewald (SPME) technique with tin-foil boundary conditions. Lennard-Jones interactions are cut off at a distance $r_{\rm cut}=\qty{0.9}{\nano\meter}$. 
The potential is shifted to zero at the cut-off separation. 
All systems are coupled to a heat bath at $\qty{300}{\kelvin}$ using v-rescale thermostat with a time constant of $\qty{0.5}{\pico\second}$.
We use MDAnalysis to treat the trajectories.

After creating the system, we first perform energy minimization. 
We then equilibrate the system in the NVT ensemble for $\qty{200}{\pico\second}$, and then in the NPT ensemble for another $\qty{200}{\pico\second}$ using a Berendsen barostat at $\qty{1}{\bar}$.
Production runs are performed in the NVT ensemble.
The integration time step is set to $\Delta t=\qty{2}{\femto\second}$.

\subsection{Permittivity of pure water under electrostatic field}
\label{app:md_perm}

We simulate a cubic box of pure water of side size $L=\qty{6.5}{\nano\meter}$.
We use the SPC/E model for water \cite{Berendsen1987Missing, Smith1993Viscosity}: a three-point charge, and one Lennard-Jones reference site model. 
The Lennard-Jones (LJ) center is placed on the oxygen. Charges are placed on the hydrogen atoms and the oxygen.

During the production run, we apply a static electrostatic field along the $x$-axis to the system using the option ``electric-field-x: $E$ 0 0 0''.
We perform 11 production runs of $\qty{20}{\nano\second}$ varying the amplitude of $E$ from $0$ to $\qty{1}{\vpnm}$ with a step of $\qty{0.1}{\vpnm}$.

In an external field, the permittivity of water is an anisotropic, diagonal tensor characterized by a component $\epsilon_\parallel$, aligned with the field direction $x$,  and a perpendicular component $\epsilon_\perp$ in the  $(y,z)$ plane.
  We calculate this tensor from the total system dipole moment~\cite{neuman83} $\MM$ that we compute for each frame after dropping the first nanosecond of the run. It is defined as  the volume integral of the polarization $\bm{\mathcal{P}}$, $\MM=\int_V \bm{\mathcal{P}}(\vec{r})\dd\rr$, where $V$ is the volume of the box. 
The permittivity obeys:
\begin{align}
\label{eps_para}
\epsilon_\parallel  &= \frac{\left\langle M_x^2 \right\rangle - \left\langle M_x \right\rangle^2}{\epsilon_0 k_B T V}+1,\\
\label{eps_perp}
\epsilon_\perp  &= \frac{\left\langle M_y^2\right\rangle + \left\langle M_z^2 \right\rangle - \left\langle M_y \right\rangle^2-\left\langle M_z \right\rangle^2}{2 \epsilon_0 k_B T V}+1.
\end{align}

\subsection{Field-dependent conductivity of aqueous electrolyte solution}
\label{app:md_conductivity}

We simulate a cubic aqueous electrolyte box of size $L=\qty{6.5}{\nano\meter}$. 
The $\qty{150}{\mmol}$ solution contains $8875$ water molecules and $25$ ion pairs of (Na$^+$, Cl$^-$), the $\qty{75}{\mmol}$ solution contains $8901$ water molecules and $12$ ion pairs. 
The Debye length of the electrolyte is equal to $\qty{0.7}{\nm}$, respectively  $\qty{1.1}{\nm}$, which is significantly smaller than the box size.
We take the following LJ parameters ($\sigma_\textrm{Na}=\qty{0.231}{\nano\meter}$, $\epsilon_\textrm{Na}=\qty{0.45}{\kilo\joule\per\mole}$) and 
($\sigma_\textrm{Cl}=\qty{0.43}{\nano\meter}$, $\epsilon_\textrm{Cl}=\qty{0.42}{\kilo\joule\per\mole}$)~\cite{Loche2021Transferable} and we use the Lorentz-Berthelot mixing rules for the LJ interactions.

For the first set of simulations, we proceed as for pure water and apply a static electrostatic field along the $x$-axis during the production run.
We perform 9 runs for which we vary the amplitude of $E$ from $\qtyrange{0.025}{0.7}{\vpnm}$.

For the second set of simulations, only the ions (Na$^+$, Cl$^-$) feel the electrostatic field, and not the water charges.
We impose an acceleration $a_\textrm{Na}=q E/m_\textrm{Na}$ to the cations and $a_\textrm{Cl}=-q E/m_\textrm{Cl}$ to the anions along the $x$ axis, where $q$ is the elementary charge and $m_i$ is the molar mass of the species $i$. 
It corresponds to the acceleration induced by an electrostatic field of amplitude $E$.
We take $m_\textrm{Na}=\qty{22.990}{\gram\per\mole}$ and $m_\textrm{Cl}=\qty{35.450}{\gram\per\mole}$.

For the two sets of simulations, we perform runs of  $\qty{200}{\nano\second}$ for the two lowest fields ($E=\qty{0.025}{\volt\per\nano\meter}$ and $E=\qty{0.05}{\volt\per\nano\meter}$), and of $\qty{20}{\nano\second}$ for the higher fields.

For each run, we measure the average velocity of the anions $v_\textrm{cl}$ and cations $v_\textrm{Na}$ after dropping the first nanosecond of the run  and we compute the conductivity with
\begin{equation}
\label{MDconductivity}
\sigma(E)=\frac{q\rho(v_\textrm{Na}-v_\textrm{Cl})}{E}.
\end{equation}
with $\rho$ the ionic density.

\subsection{Estimation of the ion mobility}
\label{app:md_ionmob}

We use a very dilute regime to compute the mobility of the ions.
We simulate a cubic box of aqueous electrolyte of side $L=\qty{10}{\nano\meter}$. The $\qty{10}{\mmol}$ solution contains 6 pairs of (Na$^+$, Cl$^-$) and 32761 water molecules. After equilibration following the procedure described in App.~\ref{app:md_methods}, we perform a simulation run of $\qty{40}{\nano\second}$ and evaluate the diffusion coefficient of the ions from the mean square displacement using the command ``gmx msd''.
We obtain $D_\textrm{Na}=\qty{1.2813\pm 0.3351e-9}{\meter^2\per\second}$ and  $D_\textrm{Cl}=\qty{2.0514\pm 0.2811e-9}{\meter^2\per\second}$.
We calculate mobilities using the Einstein relation $\kappa_i=D_i/k_BT$:
\begin{align}
\kappa_\textrm{Na}&=\qty{3.0935\pm 0.8090e11}{\meter\per\second\per\newton},\\ \kappa_\textrm{Cl}&=\qty{4.9527\pm 0.6787e11}{\meter\per\second\per\newton}.
\end{align}

\subsection{Statistical treatment}
\label{app:md_stat}

For the parallel and perpendicular permittivity (Eqs.~(\ref{eps_para},\ref{eps_perp})) and for the conductivity (Eq.~(\ref{MDconductivity})), we compute the error bars with the reblocking method~\cite{Flyvbjerg1989Error}.

\section{Anisotropic permittivity for dipoles on a lattice}
\label{app:aniso}

Here we derive the anisotropic permittivity (Eqs.~(\ref{eq:eps_para},\ref{eq:eps_perp})) from the free energy density (\ref{eq:free_energy_density}) for dipoles on a lattice.

The permittivity tensor is given by $\epsilon_{\mu\nu}=-\partial^2f(E)/(\partial E_\mu\partial E_\nu)$.
Using the fact that the free energy density $f(E)$ depends only on the magnitude of the electric field, and denoting its derivatives with primes, we find
\begin{equation}
  \epsilon_{\mu\nu}=-\delta_{\mu\nu}\frac{f'(E)}{E}+\frac{E_\mu E_\nu}{E^2}\left[\frac{f'(E)}{E}-f''(E)\right].
\end{equation}
Along the field $E_\mu E_\nu=E^2$, while transverse to it $E_\mu E_\nu=0$, which leads to
\begin{align}
  \epsilon_\parallel & = -f''(E), \label{eq:eps_para_fed}\\
  \epsilon_\perp     & = -\frac{f'(E)}{E}. \label{eq:eps_perp_fed}
\end{align}

The derivatives of the free energy density are given by
\begin{align}
  f'(E) & = -\epsilon_0 E -\rho_d p\left[\coth(x)-\frac{1}{x}\right],\\
  f''(E) & = -\epsilon_0-\beta\rho_d p^2\left[\frac{1}{x^2}-\frac{1}{\sinh(x)^2}\right].
\end{align}
Inserting these derivatives in Eqs.~(\ref{eq:eps_para_fed}, \ref{eq:eps_perp_fed}) leads to the expressions~(\ref{eq:eps_para},\ref{eq:eps_perp}).

\begin{widetext}

\section{Correlations from SDFT}
\label{app:correl}

Ionic correlations have been computed for arbitrary interactions in Ref.~\onlinecite{Demery2016Conductivity}.
The operators $\tilde R(\kk)$ and $\tilde A(\kk)$ of Eqs.~(15, 16) in Ref.~\onlinecite{Demery2016Conductivity} are given by
\begin{align}
\tilde R(\kk) & =  k^2 \begin{pmatrix}
\overline\rho_+\kappa_+ & 0 \\ 0 & \overline\rho_-\kappa_-
\end{pmatrix}
= k^2\begin{pmatrix}
r_+ & 0 \\ 0 & r_-
\end{pmatrix}, \label{eq:matrix_R}\\
\tilde A(\kk) &= \beta^{-1} \begin{pmatrix}
\frac{1}{\overline\rho_+}\left(1+i \frac{\beta z_+ q \EE\cdot\kk}{k^2} \right) + \beta\tilde V_{++}(\kk) & \beta\tilde V_{+-}(\kk) \\ \beta\tilde V_{+-}(\kk) & \frac{1}{\overline\rho_-}\left(1+i \frac{\beta z_- q \EE\cdot\kk}{k^2} \right) + \beta \tilde V_{--}(\kk)
\end{pmatrix}= \beta^{-1}\begin{pmatrix}
a & b \\ b & c
\end{pmatrix} \label{eq:matrix_A_Vab},
\end{align}
with $\bar\rho_i$ the mean density of ions of type $i$.
We still use electrostatic interactions:
\begin{equation}
\tilde V_{\alpha\beta}(\kk) = q^2 z_\alpha z_\beta \tilde G_0(\kk),
\end{equation}
where here, with $\epsilon=\epsilon_0\epsilon_r$,
\begin{equation}\label{eq:green}
\tilde G_0(\kk) = \frac{1}{\epsilon(\alpha_\parallel k_\parallel^2+\alpha_\perp k_\perp^2)}
= \frac{\tilde g (\kk)}{\epsilon}.
\end{equation}
The general form of the correlation $\tilde C(\kk)$ is given in Eq.~(24) of Ref.~\onlinecite{Demery2016Conductivity}:
\begin{multline}\label{eq:correl_val}
\tilde C = \frac{2}{(a+a^*)(c+c^*)|r_+a+r_-c^*|^2-b^2 \left[r_+(a+a^*)+r_-(c+c^*) \right]^2}\times \\
\begin{pmatrix} (c+c^*)|r_+a+r_-c^*|^2 & -b (r_+a^*+r_-c) \left[r_+(a+a^*)+r_-(c+c^*) \right] \\ 
-b (r_+a+r_-c^*) \left[r_+(a+a^*)+r_-(c+c^*) \right] & (a+a^*)|r_+a+r_-c^*|^2
\end{pmatrix}.
\end{multline}

With the notations
\begin{align}
\kappa & = \kappa_++\kappa_-,\\
\kappa' & = \kappa_+z_+-\kappa_-z_-,\\
m_\alpha^2 & = \frac{\beta q^2z_\alpha^2\bar\rho_\alpha}{\epsilon},\\
m^2 & = m_+^2+m_-^2,\\
m'^2 & = \frac{\kappa_+m_+^2+\kappa_-m_-^2}{\kappa_++\kappa_-}=\frac{\beta\sigma_0}{\epsilon (\kappa_++\kappa_-)},\\
F & = \frac{\beta qE}{m} = \frac{E}{\Ec},
\end{align}
the correlation reads
\begin{multline}\label{eq:correl_gen}
\tilde C = \left[\kappa^2(1+m^2 \tilde g)(1+m'^2 \tilde g)^2+\kappa'^2m^2(1+m_+^2\tilde g)(1+m_-^2\tilde g) \left(\frac{F k_\parallel}{k^2} \right)^2 \right]^{-1}\times\\
\begin{pmatrix}
\bar\rho_+(1+m_-^2\tilde g) \left[\kappa^2(1+m'^2\tilde g)^2+\kappa'^2m^2\left(\frac{F k_\parallel}{k^2} \right)^2 \right] & 
\sqrt{\bar\rho_+\bar\rho_-}\kappa m_+m_-\tilde g(1+m'^2\tilde g) \left[\kappa(1+m'^2\tilde g)-i\kappa'm \frac{F k_\parallel}{k^2} \right] \\ 
\sqrt{\bar\rho_+\bar\rho_-}\kappa m_+m_-\tilde g(1+m'^2\tilde g) \left[\kappa(1+m'^2\tilde g)+i\kappa'm \frac{F k_\parallel}{k^2} \right] & 
\bar\rho_-(1+m_+^2\tilde g) \left[\kappa^2(1+m'^2\tilde g)^2+\kappa'^2m^2\left(\frac{F k_\parallel}{k^2} \right)^2 \right] 
\end{pmatrix}
\end{multline}

\section{Electrostatic correction to the conductivity}
\label{app:elec}

\subsection{General case}
\label{app:elec_gen}

Using Eq.~(25) in Ref.~\onlinecite{Demery2016Conductivity}, we find the electrostatic correction to the conductivity in a $d$-dimensional space:
\begin{align}
\frac{\Delta\sigma}{\sigma_0} &= \frac{q^3}{\sigma_0 \epsilon E}\sum_{\alpha,\beta}\kappa_\alpha z_\alpha^2 z_\beta \int ik_\parallel \tilde g \tilde C_{\alpha\beta}(\kk) \frac{\dd\kk}{(2\pi)^d}\\
&= -\frac{\beta q^2\kappa'^2 m_+^2m_-^2}{\epsilon m'^2}
\int \frac{k_\parallel^2\tilde g^2(1+m'^2\tilde g)}{k^2 \left[\kappa^2(1+m^2 \tilde g)(1+m'^2 \tilde g)^2+\kappa'^2m^2(1+m_+^2\tilde g)(1+m_-^2\tilde g) \left(\frac{Fk_\parallel}{k^2} \right)^2 \right]}\frac{\dd\kk}{(2\pi)^d}.
\end{align}

To integrate the correction numerically, we specify the calculation to $d=3$ and we change the integration variable from $\kk$ to $u\in[0,\infty)$ and $y\in[-1,1]$ with $\kk=m\uu$ and $y=u_\parallel/u$, $\dd\kk= 2\pi m^3 u^2\dd u\dd y$.
The Green function reads $\tilde g=(m^2 u^2 \alpha_y)^{-1}$ (Eq.~\eqref{eq:green}) with $\alpha_y = [\epsilon_\parallel y^2+\epsilon_\perp (1-y^2)]/\epsilon$, and the correction is given by
\begin{align}
\frac{\Delta\sigma}{\sigma_0} &= -\frac{\beta q^2 m_+^2m_-^2 m^3}{(2\pi)^2\epsilon m'^2}
\int_0^\infty\dd u\int_{-1}^1\dd y \frac{y^2 u^2\tilde g^2(1+m'^2\tilde g)}{\frac{\kappa^2}{\kappa'^2}(1+m^2 \tilde g)(1+m'^2 \tilde g)^2+(1+m_+^2\tilde g)(1+m_-^2\tilde g) \frac{F^2 y^2}{u^2}} \\
& = -\frac{\beta q^2 m \mu_+^2\mu_-^2}{(2\pi)^2\epsilon \mu'^2}
\int_0^\infty\dd u\int_{-1}^1\dd y \frac{y^2\left(1+\frac{\mu'^2}{u^2\alpha_y}\right)}{\alpha_y^2 u^2\left[\frac{\kappa^2}{\kappa'^2}\left(1+\frac{1}{u^2\alpha_y}\right)\left(1+\frac{\mu'^2}{u^2\alpha_y}\right)^2+\left(1+\frac{\mu_+^2}{u^2\alpha_y}\right)\left(1+\frac{\mu_-^2}{u^2\alpha_y}\right) \frac{F^2 y^2}{u^2}\right]}\\
& = -\frac{\beta q^2 m \mu_+^2\mu_-^2}{2\pi^2\epsilon \mu'^2}
\int_0^\infty\dd u\int_0^1\dd y \frac{y^2 u^2\left(\alpha_yu^2+\mu'^2\right)}{\frac{\kappa^2}{\kappa'^2}\left(\alpha_yu^2+1\right)\left(\alpha_yu^2+\mu'^2\right)^2+F^2 y^2\alpha_y\left(\alpha_yu^2+\mu_+^2\right)\left(\alpha_yu^2+\mu_-^2\right) }.
\end{align}
We have introduced $\mu'=m'/m$, $\mu_\pm=m_\pm/m$, and used the fact that the integral is even in $y$.

\subsection{Monovalent salt}
\label{app:elec_mono}

For a monovalent salt, $z_+=-z_-=1$ and $\rho_+=\rho_-$, so that $m_i^2=m'^2=m^2/2$, $\kappa'=\kappa$, $\mu_i=\mu'=1/\sqrt{2}$.
The expression above simplifies to
\begin{equation}
  \frac{\Delta\sigma_\mathrm{el}}{\sigma_0}
  = -\frac{\beta q^2 m }{2\pi^2\epsilon}
\int_0^\infty\dd u\int_0^1\dd y \frac{y^2 u^2}{(2u^2\alpha_y+1)(u^2\alpha_y+1+F^2 y^2\alpha_y)}
  = -\frac{\beta q^2 m }{4\sqrt{2}\pi\epsilon_0\epsilon_r}
\int_0^1 \frac{y^2 \dd y}{\alpha_y^{3/2}\left[1+\sqrt{2(1+F^2y^2\alpha_y)}\right]}.
\end{equation}

\subsection{Small field expansion}
\label{app:elec_small}

We write the electrostatic correction as
\begin{equation}
  \frac{\dse}{\sigma_0} = -\frac{\beta q^2 m \mu_+^2\mu_-^2}{2\pi^2\epsilon \mu'^2} s_e,
\end{equation}
with 
\begin{equation}
s_e = \int_0^\infty\dd u\int_0^1\dd y \frac{y^2 u^2\left(\alpha_yu^2+\mu'^2\right)}{\frac{\kappa^2}{\kappa'^2}\left(\alpha_yu^2+1\right)\left(\alpha_yu^2+\mu'^2\right)^2+F^2 y^2\alpha_y\left(\alpha_yu^2+\mu_+^2\right)\left(\alpha_yu^2+\mu_-^2\right) }.  
\end{equation}

After a few lines of calculation, we get at order $E^2$
\begin{multline}
s_e = \frac{\kappa'^2}{3\kappa^2}\int_0^\infty\dd u\frac{u^2}{(1+u^2)(u^2+\mu'^2)}
-\frac{\kappa'^4}{5\kappa^4} F^2\int_0^\infty\dd u\frac{u^2(u^2+\mu_+^2)(u^2+\mu_-^2)}{(u^2+1)^2(u^2+\mu'^2)^3}\\+\frac{11}{225}\frac{\kappa'^2}{\kappa^2}\frac{\epsilon_r-1}{\epsilon_r} x^2\int_0^\infty\dd u\frac{u^4(2u^2+1+\mu'^2)}{(u^2+1)^2(u^2+\mu'^2)^2},
\end{multline}
where $F=E/\Ec$ and $x=E/\Ew$.

\section{Hydrodynamic correction to the conductivity}\label{}
\label{app:hydro}

\subsection{General case}
\label{app:hydro_gen}

The hydrodynamic correction to the conductivity is~\cite{Bonneau2023Temporal} (Eq.~(17)):
\begin{equation}
\frac{\Delta\sigma\ind{hyd}}{\sigma_0}=q^2\int\tilde \mcO_{11}(\kk) \sum_{\alpha\beta} z_\alpha z_\beta\left[\tilde C_{\alpha\beta}(\kk)-\bar\rho_\alpha\delta_{\alpha\beta}\right] \frac{\dd\kk}{(2\pi)^d}.
\end{equation}
Using the correlation (\ref{eq:correl_gen}), we find
\begin{multline}
\frac{\Delta\sigma\ind{hyd}}{\sigma_0}
=-\frac{q^2}{\eta\sigma_0} \int \frac{\dd\kk}{(2\pi)^d} \frac{k_\perp^2}{k^4}\tilde g\\\times \frac{\kappa^2\left(1+m'^2\tilde g\right)^2\left(z_+\sqrt{\bar\rho_+}m_++z_-\sqrt{\bar\rho_-}m_-\right)^2+\kappa'^2m^2\left[z_+^2\bar\rho_+m_+^2(1+m_-^2\tilde g)+z_-^2\bar\rho_-m_-^2(1+m^2_+)\tilde g\right] \left(\frac{Fk_\parallel}{k^2} \right)^2}{\kappa^2(1+m^2\tilde g)(1+m'^2\tilde g)^2+\kappa'^2m^2(1+m_+^2\tilde g)(1+m_-^2\tilde g) \left(\frac{Fk_\parallel}{k^2} \right)^2 } .
\end{multline}

To perform the numerical integration, we follow the same steps as for the electrostatic correction:
\begin{align}
\frac{\Delta\sigma\ind{hyd}}{\sigma_0}
& =-\frac{q^2 m}{(2\pi)^2 \eta\sigma_0} \int_0^\infty\dd u\int_{-1}^1\dd y (1-y^2)\tilde g \nonumber\\
&\qquad\times \frac{\frac{\kappa^2}{\kappa'^2}\left(1+m'^2\tilde g\right)^2\left(\sum_\alpha z_\alpha\sqrt{\bar\rho_\alpha}m_\alpha\right)^2+\left[z_+^2\bar\rho_+m_+^2(1+m_-^2\tilde g)+z_-^2\bar\rho_-m_-^2(1+m^2_+\tilde g)\right] \frac{F^2 y^2}{u^2}}{\frac{\kappa^2}{\kappa'^2}(1+m^2\tilde g)(1+m'^2\tilde g)^2+(1+m_+^2\tilde g)(1+m_-^2\tilde g) \frac{F^2 y^2}{u^2} }\\
& =-\frac{q^2 m}{(2\pi)^2 \eta\sigma_0} \int_0^\infty\dd u\int_{-1}^1\dd y \frac{1-y^2}{u^2\alpha_y} \nonumber\\
& \qquad \times \frac{\frac{\kappa^2}{\kappa'^2}\left(1+\frac{\mu'^2}{u^2\alpha_y}\right)^2\left(\sum_\alpha z_\alpha\sqrt{\bar\rho_\alpha}\mu_\alpha\right)^2+\left[z_+^2\bar\rho_+\mu_+^2(1+\frac{\mu_-^2}{u^2\alpha_y})+z_-^2\bar\rho_-\mu_-^2(1+\frac{\mu_+^2}{u^2\alpha_y})\right] \frac{F^2 y^2}{u^2}}{\frac{\kappa^2}{\kappa'^2}(1+\frac{1}{u^2\alpha_y})(1+\frac{\mu'^2}{u^2\alpha_y})^2+(1+\frac{\mu_+^2}{u^2\alpha_y})(1+\frac{\mu_-^2}{u^2\alpha_y}) \frac{F^2 y^2}{u^2} }\\
& =-\frac{q^2 m}{2\pi^2 \eta\sigma_0} \int_0^\infty\dd u\int_0^1\dd y (1-y^2) \nonumber\\
& \qquad\times \frac{\frac{\kappa^2}{\kappa'^2}\left(u^2\alpha_y+\mu'^2\right)^2\left(\sum_\alpha z_\alpha\sqrt{\bar\rho_\alpha}\mu_\alpha\right)^2+F^2 y^2\alpha_y\left[z_+^2\bar\rho_+\mu_+^2(u^2\alpha_y+\mu_-^2)+z_-^2\bar\rho_-\mu_-^2(u^2\alpha_y+\mu_+^2)\right]}
{\frac{\kappa^2}{\kappa'^2}(u^2\alpha_y+1)(u^2\alpha_y+\mu'^2)^2+F^2 y^2\alpha_y(u^2\alpha_y+\mu_+^2)(u^2\alpha_y+\mu_-^2) }.
\end{align}

\subsection{Monovalent salt}
\label{app:hydro_mono}

For a monovalent salt the expression above simplifies into
\begin{equation}
\frac{\Delta\sigma\ind{hyd}}{\sigma_0}
=-\frac{m}{\pi^2 \eta\kappa} \int_0^\infty\dd u\int_0^1\dd y \frac{(1-y^2) (2u^2\alpha_y+1+F^2 y^2\alpha_y)}{(2u^2\alpha_y+1)(u^2\alpha_y+1+F^2 y^2\alpha_y)}
=-\frac{m}{2\sqrt{2}\pi \eta\kappa} \int_0^1\dd y\frac{1-y^2}{\sqrt{\alpha_y}}\left[1+\frac{1}{1+\sqrt{2(1+F^2 y^2\alpha_y)}}\right].
\end{equation}

\subsection{Small field expansion}
\label{app:hydro_small}

For the hydrodynamic correction, we denote
\begin{equation}
  \frac{\dsh}{\sigma_0} = -\frac{q^2 m}{2\pi^2 \eta\sigma_0} s_h,
\end{equation}
where
\begin{equation}
  s_h
=\int_0^\infty\dd u\int_0^1\dd y (1-y^2) \frac{\frac{\kappa^2}{\kappa'^2}\left(u^2\alpha_y+\mu'^2\right)^2\left(\sum_i z_i\sqrt{\bar\rho_i}\mu_i\right)^2+F^2 y^2\alpha_y\left[z_+^2\bar\rho_+\mu_+^2(u^2\alpha_y+\mu_-^2)+z_-^2\bar\rho_-\mu_-^2(u^2\alpha_y+\mu_+^2)\right]}
{\frac{\kappa^2}{\kappa'^2}(u^2\alpha_y+1)(u^2\alpha_y+\mu'^2)^2+F^2 y^2\alpha_y(u^2\alpha_y+\mu_+^2)(u^2\alpha_y+\mu_-^2) }.
\end{equation}

Using the same approach, we find at order $E^2$
\begin{multline}
s_h
=\frac{\pi}{3}\left(\sum_i \sqrt{z_i^2\mu_i^2\bar\rho_i}\right)^2+\frac{7\pi}{450}\frac{\epsilon_r-1}{\epsilon_r}\left(\sum_i \sqrt{z_i^2\mu_i^2\bar\rho_i}\right)^2 x^2
\\+\frac{2}{15}\frac{\kappa'^2}{\kappa^2}F^2\left[\int_0^\infty\dd u \frac{z_+^2\bar\rho_+\mu_+^2(u^2+\mu_-^2)+z_-^2\bar\rho_-\mu_-^2(u^2+\mu_+^2)}{(u^2+1)(u^2+\mu'^2)^2}
-\left(\sum_i \sqrt{z_i^2\mu_i^2\bar\rho_i}\right)^2\int_0^\infty\dd u \frac{(u^2+\mu_+^2)(u^2+\mu_-^2)}{(u^2+1)^2(u^2+\mu'^2)^2}\right],
\end{multline}
where $F=E/\Ec$ and $x=E/\Ew$.

\end{widetext}


\begin{thebibliography}{52}%
\makeatletter
\providecommand \@ifxundefined [1]{%
 \@ifx{#1\undefined}
}%
\providecommand \@ifnum [1]{%
 \ifnum #1\expandafter \@firstoftwo
 \else \expandafter \@secondoftwo
 \fi
}%
\providecommand \@ifx [1]{%
 \ifx #1\expandafter \@firstoftwo
 \else \expandafter \@secondoftwo
 \fi
}%
\providecommand \natexlab [1]{#1}%
\providecommand \enquote  [1]{``#1''}%
\providecommand \bibnamefont  [1]{#1}%
\providecommand \bibfnamefont [1]{#1}%
\providecommand \citenamefont [1]{#1}%
\providecommand \href@noop [0]{\@secondoftwo}%
\providecommand \href [0]{\begingroup \@sanitize@url \@href}%
\providecommand \@href[1]{\@@startlink{#1}\@@href}%
\providecommand \@@href[1]{\endgroup#1\@@endlink}%
\providecommand \@sanitize@url [0]{\catcode `\\12\catcode `\$12\catcode
  `\&12\catcode `\#12\catcode `\^12\catcode `\_12\catcode `\%12\relax}%
\providecommand \@@startlink[1]{}%
\providecommand \@@endlink[0]{}%
\providecommand \url  [0]{\begingroup\@sanitize@url \@url }%
\providecommand \@url [1]{\endgroup\@href {#1}{\urlprefix }}%
\providecommand \urlprefix  [0]{URL }%
\providecommand \Eprint [0]{\href }%
\providecommand \doibase [0]{http://dx.doi.org/}%
\providecommand \selectlanguage [0]{\@gobble}%
\providecommand \bibinfo  [0]{\@secondoftwo}%
\providecommand \bibfield  [0]{\@secondoftwo}%
\providecommand \translation [1]{[#1]}%
\providecommand \BibitemOpen [0]{}%
\providecommand \bibitemStop [0]{}%
\providecommand \bibitemNoStop [0]{.\EOS\space}%
\providecommand \EOS [0]{\spacefactor3000\relax}%
\providecommand \BibitemShut  [1]{\csname bibitem#1\endcsname}%
\let\auto@bib@innerbib\@empty
\bibitem [{\citenamefont {Eckstrom}\ and\ \citenamefont
  {Schmelzer}(1939)}]{Eckstrom1939Wien}%
  \BibitemOpen
  \bibfield  {author} {\bibinfo {author} {\bibfnamefont {H.~C.}\ \bibnamefont
  {Eckstrom}}\ and\ \bibinfo {author} {\bibfnamefont {C.}~\bibnamefont
  {Schmelzer}},\ }\bibfield  {title} {\enquote {\bibinfo {title} {The wien
  effect: Deviations of electrolytic solutions from ohm's law under high field
  strengths.}}\ }\href {\doibase 10.1021/cr60079a001} {\bibfield  {journal}
  {\bibinfo  {journal} {Chemical Reviews}\ }\textbf {\bibinfo {volume} {24}},\
  \bibinfo {pages} {367--414} (\bibinfo {year} {1939})}\BibitemShut {NoStop}%
\bibitem [{\citenamefont {Wilson}(1936)}]{Wilson1936}%
  \BibitemOpen
  \bibfield  {author} {\bibinfo {author} {\bibfnamefont {W.~S.}\ \bibnamefont
  {Wilson}},\ }\emph {\bibinfo {title} {The theory of the Wien effect for a
  binary electrolyte}},\ \href@noop {} {Ph.D. thesis},\ \bibinfo  {school}
  {Yale University} (\bibinfo {year} {1936})\BibitemShut {NoStop}%
\bibitem [{\citenamefont {Debye}\ and\ \citenamefont
  {Hückel}(1923)}]{Debye1923}%
  \BibitemOpen
  \bibfield  {author} {\bibinfo {author} {\bibfnamefont {P.}~\bibnamefont
  {Debye}}\ and\ \bibinfo {author} {\bibfnamefont {E.}~\bibnamefont
  {Hückel}},\ }\bibfield  {title} {\enquote {\bibinfo {title} {{Theory of
  electrolytes—part II: law of the limit of electrolytic conduction}},}\
  }\href@noop {} {\bibfield  {journal} {\bibinfo  {journal} {{Physikalische
  Zeitschrift}}\ }\textbf {\bibinfo {volume} {24}},\ \bibinfo {pages}
  {305--325} (\bibinfo {year} {1923})}\BibitemShut {NoStop}%
\bibitem [{\citenamefont {Onsager}(1927)}]{Onsager1927}%
  \BibitemOpen
  \bibfield  {author} {\bibinfo {author} {\bibfnamefont {L.}~\bibnamefont
  {Onsager}},\ }\bibfield  {title} {\enquote {\bibinfo {title} {{Report on a
  revision of the conductivity theory}},}\ }\href {\doibase
  10.1039/tf9272300341} {\bibfield  {journal} {\bibinfo  {journal}
  {{Transactions of the Faraday Society}}\ }\textbf {\bibinfo {volume} {23}},\
  \bibinfo {pages} {341--349} (\bibinfo {year} {1927})}\BibitemShut {NoStop}%
\bibitem [{\citenamefont {Onsager}\ and\ \citenamefont
  {Kim}(1957)}]{Onsager1957Wien}%
  \BibitemOpen
  \bibfield  {author} {\bibinfo {author} {\bibfnamefont {L.}~\bibnamefont
  {Onsager}}\ and\ \bibinfo {author} {\bibfnamefont {S.~K.}\ \bibnamefont
  {Kim}},\ }\bibfield  {title} {\enquote {\bibinfo {title} {{Wien Effect in
  Simple Strong Electrolytes}},}\ }\href {\doibase 10.1021/j150548a015}
  {\bibfield  {journal} {\bibinfo  {journal} {{The Journal of Physical
  Chemistry}}\ }\textbf {\bibinfo {volume} {61}},\ \bibinfo {pages} {198--215}
  (\bibinfo {year} {1957})}\BibitemShut {NoStop}%
\bibitem [{\citenamefont {Onsager}(1934)}]{Onsager1934}%
  \BibitemOpen
  \bibfield  {author} {\bibinfo {author} {\bibfnamefont {L.}~\bibnamefont
  {Onsager}},\ }\bibfield  {title} {\enquote {\bibinfo {title} {{Deviations
  from Ohm's Law in Weak Electrolytes}},}\ }\href {\doibase
  http://dx.doi.org/10.1063/1.1749541} {\bibfield  {journal} {\bibinfo
  {journal} {{The Journal of Chemical Physics}}\ }\textbf {\bibinfo {volume}
  {2}},\ \bibinfo {pages} {599--615} (\bibinfo {year} {1934})}\BibitemShut
  {NoStop}%
\bibitem [{\citenamefont {Feng}\ \emph {et~al.}(2016)\citenamefont {Feng},
  \citenamefont {Liu}, \citenamefont {Graf}, \citenamefont {Dumcenco},
  \citenamefont {Kis}, \citenamefont {Di~Ventra},\ and\ \citenamefont
  {Radenovic}}]{Feng2016Observation}%
  \BibitemOpen
  \bibfield  {author} {\bibinfo {author} {\bibfnamefont {J.}~\bibnamefont
  {Feng}}, \bibinfo {author} {\bibfnamefont {K.}~\bibnamefont {Liu}}, \bibinfo
  {author} {\bibfnamefont {M.}~\bibnamefont {Graf}}, \bibinfo {author}
  {\bibfnamefont {D.}~\bibnamefont {Dumcenco}}, \bibinfo {author}
  {\bibfnamefont {A.}~\bibnamefont {Kis}}, \bibinfo {author} {\bibfnamefont
  {M.}~\bibnamefont {Di~Ventra}}, \ and\ \bibinfo {author} {\bibfnamefont
  {A.}~\bibnamefont {Radenovic}},\ }\bibfield  {title} {\enquote {\bibinfo
  {title} {Observation of ionic coulomb blockade in nanopores},}\ }\href
  {\doibase 10.1038/nmat4607} {\bibfield  {journal} {\bibinfo  {journal}
  {Nature Materials}\ }\textbf {\bibinfo {volume} {15}},\ \bibinfo {pages}
  {850--855} (\bibinfo {year} {2016})}\BibitemShut {NoStop}%
\bibitem [{\citenamefont {Cai}\ \emph {et~al.}(2022)\citenamefont {Cai},
  \citenamefont {Griffin}, \citenamefont {Guarochico-Moreira}, \citenamefont
  {Barry}, \citenamefont {Xin}, \citenamefont {Yagmurcukardes}, \citenamefont
  {Zhang}, \citenamefont {Geim}, \citenamefont {Peeters},\ and\ \citenamefont
  {Lozada-Hidalgo}}]{Cai2022Wien}%
  \BibitemOpen
  \bibfield  {author} {\bibinfo {author} {\bibfnamefont {J.}~\bibnamefont
  {Cai}}, \bibinfo {author} {\bibfnamefont {E.}~\bibnamefont {Griffin}},
  \bibinfo {author} {\bibfnamefont {V.~H.}\ \bibnamefont {Guarochico-Moreira}},
  \bibinfo {author} {\bibfnamefont {D.}~\bibnamefont {Barry}}, \bibinfo
  {author} {\bibfnamefont {B.}~\bibnamefont {Xin}}, \bibinfo {author}
  {\bibfnamefont {M.}~\bibnamefont {Yagmurcukardes}}, \bibinfo {author}
  {\bibfnamefont {S.}~\bibnamefont {Zhang}}, \bibinfo {author} {\bibfnamefont
  {A.~K.}\ \bibnamefont {Geim}}, \bibinfo {author} {\bibfnamefont {F.~M.}\
  \bibnamefont {Peeters}}, \ and\ \bibinfo {author} {\bibfnamefont
  {M.}~\bibnamefont {Lozada-Hidalgo}},\ }\bibfield  {title} {\enquote {\bibinfo
  {title} {Wien effect in interfacial water dissociation through
  proton-permeable graphene electrodes},}\ }\href {\doibase
  10.1038/s41467-022-33451-1} {\bibfield  {journal} {\bibinfo  {journal}
  {Nature Communications}\ }\textbf {\bibinfo {volume} {13}},\ \bibinfo {pages}
  {5776} (\bibinfo {year} {2022})}\BibitemShut {NoStop}%
\bibitem [{\citenamefont {Montenegro}\ \emph {et~al.}(2021)\citenamefont
  {Montenegro}, \citenamefont {Dutta}, \citenamefont {Mammetkuliev},
  \citenamefont {Shi}, \citenamefont {Hou}, \citenamefont {Bhattacharyya},
  \citenamefont {Zhao}, \citenamefont {Cronin},\ and\ \citenamefont
  {Benderskii}}]{Montenegro2021}%
  \BibitemOpen
  \bibfield  {author} {\bibinfo {author} {\bibfnamefont {A.}~\bibnamefont
  {Montenegro}}, \bibinfo {author} {\bibfnamefont {C.}~\bibnamefont {Dutta}},
  \bibinfo {author} {\bibfnamefont {M.}~\bibnamefont {Mammetkuliev}}, \bibinfo
  {author} {\bibfnamefont {H.}~\bibnamefont {Shi}}, \bibinfo {author}
  {\bibfnamefont {B.}~\bibnamefont {Hou}}, \bibinfo {author} {\bibfnamefont
  {D.}~\bibnamefont {Bhattacharyya}}, \bibinfo {author} {\bibfnamefont
  {B.}~\bibnamefont {Zhao}}, \bibinfo {author} {\bibfnamefont {S.~B.}\
  \bibnamefont {Cronin}}, \ and\ \bibinfo {author} {\bibfnamefont {A.~V.}\
  \bibnamefont {Benderskii}},\ }\bibfield  {title} {\enquote {\bibinfo {title}
  {Asymmetric response of interfacial water to applied electric fields},}\
  }\href {\doibase 10.1038/s41586-021-03504-4} {\bibfield  {journal} {\bibinfo
  {journal} {Nature}\ }\textbf {\bibinfo {volume} {594}},\ \bibinfo {pages}
  {62--65} (\bibinfo {year} {2021})}\BibitemShut {NoStop}%
\bibitem [{\citenamefont {Robin}\ \emph {et~al.}(2023)\citenamefont {Robin},
  \citenamefont {Emmerich}, \citenamefont {Ismail}, \citenamefont {Niguès},
  \citenamefont {You}, \citenamefont {Nam}, \citenamefont {Keerthi},
  \citenamefont {Siria}, \citenamefont {Geim}, \citenamefont {Radha},\ and\
  \citenamefont {Bocquet}}]{Robin2023Long-term}%
  \BibitemOpen
  \bibfield  {author} {\bibinfo {author} {\bibfnamefont {P.}~\bibnamefont
  {Robin}}, \bibinfo {author} {\bibfnamefont {T.}~\bibnamefont {Emmerich}},
  \bibinfo {author} {\bibfnamefont {A.}~\bibnamefont {Ismail}}, \bibinfo
  {author} {\bibfnamefont {A.}~\bibnamefont {Niguès}}, \bibinfo {author}
  {\bibfnamefont {Y.}~\bibnamefont {You}}, \bibinfo {author} {\bibfnamefont
  {G.-H.}\ \bibnamefont {Nam}}, \bibinfo {author} {\bibfnamefont
  {A.}~\bibnamefont {Keerthi}}, \bibinfo {author} {\bibfnamefont
  {A.}~\bibnamefont {Siria}}, \bibinfo {author} {\bibfnamefont {A.~K.}\
  \bibnamefont {Geim}}, \bibinfo {author} {\bibfnamefont {B.}~\bibnamefont
  {Radha}}, \ and\ \bibinfo {author} {\bibfnamefont {L.}~\bibnamefont
  {Bocquet}},\ }\bibfield  {title} {\enquote {\bibinfo {title} {Long-term
  memory and synapse-like dynamics in two-dimensional nanofluidic channels},}\
  }\href {\doibase 10.1126/science.adc9931} {\bibfield  {journal} {\bibinfo
  {journal} {Science}\ }\textbf {\bibinfo {volume} {379}},\ \bibinfo {pages}
  {161--167} (\bibinfo {year} {2023})}\BibitemShut {NoStop}%
\bibitem [{\citenamefont {Robin}, \citenamefont {Kavokine},\ and\ \citenamefont
  {Bocquet}(2021)}]{Robin2021Modeling}%
  \BibitemOpen
  \bibfield  {author} {\bibinfo {author} {\bibfnamefont {P.}~\bibnamefont
  {Robin}}, \bibinfo {author} {\bibfnamefont {N.}~\bibnamefont {Kavokine}}, \
  and\ \bibinfo {author} {\bibfnamefont {L.}~\bibnamefont {Bocquet}},\
  }\bibfield  {title} {\enquote {\bibinfo {title} {Modeling of emergent memory
  and voltage spiking in ionic transport through angstrom-scale slits},}\
  }\href {\doibase 10.1126/science.abf7923} {\bibfield  {journal} {\bibinfo
  {journal} {Science}\ }\textbf {\bibinfo {volume} {373}},\ \bibinfo {pages}
  {687--691} (\bibinfo {year} {2021})}\BibitemShut {NoStop}%
\bibitem [{\citenamefont {Lesnicki}\ \emph {et~al.}(2020)\citenamefont
  {Lesnicki}, \citenamefont {Gao}, \citenamefont {Rotenberg},\ and\
  \citenamefont {Limmer}}]{Lesnicki2020}%
  \BibitemOpen
  \bibfield  {author} {\bibinfo {author} {\bibfnamefont {D.}~\bibnamefont
  {Lesnicki}}, \bibinfo {author} {\bibfnamefont {C.~Y.}\ \bibnamefont {Gao}},
  \bibinfo {author} {\bibfnamefont {B.}~\bibnamefont {Rotenberg}}, \ and\
  \bibinfo {author} {\bibfnamefont {D.~T.}\ \bibnamefont {Limmer}},\ }\bibfield
   {title} {\enquote {\bibinfo {title} {Field-dependent ionic conductivities
  from generalized fluctuation-dissipation relations},}\ }\href {\doibase
  10.1103/PhysRevLett.124.206001} {\bibfield  {journal} {\bibinfo  {journal}
  {Phys. Rev. Lett.}\ }\textbf {\bibinfo {volume} {124}},\ \bibinfo {pages}
  {206001} (\bibinfo {year} {2020})}\BibitemShut {NoStop}%
\bibitem [{\citenamefont {Lesnicki}\ \emph {et~al.}(2021)\citenamefont
  {Lesnicki}, \citenamefont {Gao}, \citenamefont {Limmer},\ and\ \citenamefont
  {Rotenberg}}]{Lesnicki2021}%
  \BibitemOpen
  \bibfield  {author} {\bibinfo {author} {\bibfnamefont {D.}~\bibnamefont
  {Lesnicki}}, \bibinfo {author} {\bibfnamefont {C.~Y.}\ \bibnamefont {Gao}},
  \bibinfo {author} {\bibfnamefont {D.~T.}\ \bibnamefont {Limmer}}, \ and\
  \bibinfo {author} {\bibfnamefont {B.}~\bibnamefont {Rotenberg}},\ }\bibfield
  {title} {\enquote {\bibinfo {title} {On the molecular correlations that
  result in field-dependent conductivities in electrolyte solutions},}\ }\href
  {\doibase 10.1063/5.0052860} {\bibfield  {journal} {\bibinfo  {journal} {The
  Journal of Chemical Physics}\ }\textbf {\bibinfo {volume} {155}},\ \bibinfo
  {pages} {014507} (\bibinfo {year} {2021})}\BibitemShut {NoStop}%
\bibitem [{\citenamefont {Dean}(1996)}]{Dean1996}%
  \BibitemOpen
  \bibfield  {author} {\bibinfo {author} {\bibfnamefont {D.~S.}\ \bibnamefont
  {Dean}},\ }\bibfield  {title} {\enquote {\bibinfo {title} {{Langevin equation
  for the density of a system of interacting Langevin processes}},}\ }\href
  {\doibase 10.1088/0305-4470/29/24/001} {\bibfield  {journal} {\bibinfo
  {journal} {{Journal of Physics A: Mathematical and General}}\ }\textbf
  {\bibinfo {volume} {29}},\ \bibinfo {pages} {L613--L617} (\bibinfo {year}
  {1996})}\BibitemShut {NoStop}%
\bibitem [{\citenamefont {Démery}\ and\ \citenamefont
  {Dean}(2016)}]{Demery2016Conductivity}%
  \BibitemOpen
  \bibfield  {author} {\bibinfo {author} {\bibfnamefont {V.}~\bibnamefont
  {Démery}}\ and\ \bibinfo {author} {\bibfnamefont {D.~S.}\ \bibnamefont
  {Dean}},\ }\bibfield  {title} {\enquote {\bibinfo {title} {{The conductivity
  of strong electrolytes from stochastic density functional theory}},}\ }\href
  {\doibase 10.1088/1742-5468/2016/02/023106} {\bibfield  {journal} {\bibinfo
  {journal} {{Journal of Statistical Mechanics: Theory and Experiment}}\
  }\textbf {\bibinfo {volume} {2016}},\ \bibinfo {pages} {023106} (\bibinfo
  {year} {2016})}\BibitemShut {NoStop}%
\bibitem [{\citenamefont {Péraud}\ \emph {et~al.}(2017)\citenamefont
  {Péraud}, \citenamefont {Nonaka}, \citenamefont {Bell}, \citenamefont
  {Donev},\ and\ \citenamefont {Garcia}}]{Peraud2017}%
  \BibitemOpen
  \bibfield  {author} {\bibinfo {author} {\bibfnamefont {J.-P.}\ \bibnamefont
  {Péraud}}, \bibinfo {author} {\bibfnamefont {A.~J.}\ \bibnamefont {Nonaka}},
  \bibinfo {author} {\bibfnamefont {J.~B.}\ \bibnamefont {Bell}}, \bibinfo
  {author} {\bibfnamefont {A.}~\bibnamefont {Donev}}, \ and\ \bibinfo {author}
  {\bibfnamefont {A.~L.}\ \bibnamefont {Garcia}},\ }\bibfield  {title}
  {\enquote {\bibinfo {title} {{Fluctuation-enhanced electric conductivity in
  electrolyte solutions}},}\ }\href {\doibase 10.1073/pnas.1714464114}
  {\bibfield  {journal} {\bibinfo  {journal} {{Proceedings of the National
  Academy of Sciences}}\ }\textbf {\bibinfo {volume} {114}},\ \bibinfo {pages}
  {10829--10833} (\bibinfo {year} {2017})}\BibitemShut {NoStop}%
\bibitem [{\citenamefont {Donev}\ \emph {et~al.}(2019)\citenamefont {Donev},
  \citenamefont {Garcia}, \citenamefont {Péraud}, \citenamefont {Nonaka},\
  and\ \citenamefont {Bell}}]{Donev2019}%
  \BibitemOpen
  \bibfield  {author} {\bibinfo {author} {\bibfnamefont {A.}~\bibnamefont
  {Donev}}, \bibinfo {author} {\bibfnamefont {A.~L.}\ \bibnamefont {Garcia}},
  \bibinfo {author} {\bibfnamefont {J.-P.}\ \bibnamefont {Péraud}}, \bibinfo
  {author} {\bibfnamefont {A.~J.}\ \bibnamefont {Nonaka}}, \ and\ \bibinfo
  {author} {\bibfnamefont {J.~B.}\ \bibnamefont {Bell}},\ }\bibfield  {title}
  {\enquote {\bibinfo {title} {{Fluctuating Hydrodynamics and
  Debye-Hückel-Onsager Theory for Electrolytes}},}\ }\href {\doibase
  https://doi.org/10.1016/j.coelec.2018.09.004} {\bibfield  {journal} {\bibinfo
   {journal} {{Current Opinion in Electrochemistry}}\ }\textbf {\bibinfo
  {volume} {13}},\ \bibinfo {pages} {1--10} (\bibinfo {year}
  {2019})}\BibitemShut {NoStop}%
\bibitem [{\citenamefont {Avni}\ \emph {et~al.}(2022)\citenamefont {Avni},
  \citenamefont {Adar}, \citenamefont {Andelman},\ and\ \citenamefont
  {Orland}}]{Avni2022Conductivity}%
  \BibitemOpen
  \bibfield  {author} {\bibinfo {author} {\bibfnamefont {Y.}~\bibnamefont
  {Avni}}, \bibinfo {author} {\bibfnamefont {R.~M.}\ \bibnamefont {Adar}},
  \bibinfo {author} {\bibfnamefont {D.}~\bibnamefont {Andelman}}, \ and\
  \bibinfo {author} {\bibfnamefont {H.}~\bibnamefont {Orland}},\ }\bibfield
  {title} {\enquote {\bibinfo {title} {Conductivity of concentrated
  electrolytes},}\ }\href {\doibase 10.1103/PhysRevLett.128.098002} {\bibfield
  {journal} {\bibinfo  {journal} {Phys. Rev. Lett.}\ }\textbf {\bibinfo
  {volume} {128}},\ \bibinfo {pages} {098002} (\bibinfo {year}
  {2022})}\BibitemShut {NoStop}%
\bibitem [{\citenamefont {Avni}, \citenamefont {Andelman},\ and\ \citenamefont
  {Orland}(2022)}]{Avni2022Conductance}%
  \BibitemOpen
  \bibfield  {author} {\bibinfo {author} {\bibfnamefont {Y.}~\bibnamefont
  {Avni}}, \bibinfo {author} {\bibfnamefont {D.}~\bibnamefont {Andelman}}, \
  and\ \bibinfo {author} {\bibfnamefont {H.}~\bibnamefont {Orland}},\
  }\bibfield  {title} {\enquote {\bibinfo {title} {Conductance of concentrated
  electrolytes: Multivalency and the wien effect},}\ }\href {\doibase
  10.1063/5.0111645} {\bibfield  {journal} {\bibinfo  {journal} {The Journal of
  Chemical Physics}\ }\textbf {\bibinfo {volume} {157}},\ \bibinfo {pages}
  {154502} (\bibinfo {year} {2022})}\BibitemShut {NoStop}%
\bibitem [{\citenamefont {Bernard}\ \emph {et~al.}(2023)\citenamefont
  {Bernard}, \citenamefont {Jardat}, \citenamefont {Rotenberg},\ and\
  \citenamefont {Illien}}]{Bernard2023Analytical}%
  \BibitemOpen
  \bibfield  {author} {\bibinfo {author} {\bibfnamefont {O.}~\bibnamefont
  {Bernard}}, \bibinfo {author} {\bibfnamefont {M.}~\bibnamefont {Jardat}},
  \bibinfo {author} {\bibfnamefont {B.}~\bibnamefont {Rotenberg}}, \ and\
  \bibinfo {author} {\bibfnamefont {P.}~\bibnamefont {Illien}},\ }\bibfield
  {title} {\enquote {\bibinfo {title} {{On analytical theories for conductivity
  and self-diffusion in concentrated electrolytes}},}\ }\href {\doibase
  10.1063/5.0165533} {\bibfield  {journal} {\bibinfo  {journal} {The Journal of
  Chemical Physics}\ }\textbf {\bibinfo {volume} {159}},\ \bibinfo {pages}
  {164105} (\bibinfo {year} {2023})}\BibitemShut {NoStop}%
\bibitem [{\citenamefont {Dufrêche}\ \emph {et~al.}(2005)\citenamefont
  {Dufrêche}, \citenamefont {Bernard}, \citenamefont {Durand-Vidal},\ and\
  \citenamefont {Turq}}]{Dufreche2005Analytical}%
  \BibitemOpen
  \bibfield  {author} {\bibinfo {author} {\bibfnamefont {J.-F.}\ \bibnamefont
  {Dufrêche}}, \bibinfo {author} {\bibfnamefont {O.}~\bibnamefont {Bernard}},
  \bibinfo {author} {\bibfnamefont {S.}~\bibnamefont {Durand-Vidal}}, \ and\
  \bibinfo {author} {\bibfnamefont {P.}~\bibnamefont {Turq}},\ }\bibfield
  {title} {\enquote {\bibinfo {title} {Analytical theories of transport in
  concentrated electrolyte solutions from the msa},}\ }\href {\doibase
  10.1021/jp050387y} {\bibfield  {journal} {\bibinfo  {journal} {The Journal of
  Physical Chemistry B}\ }\textbf {\bibinfo {volume} {109}},\ \bibinfo {pages}
  {9873--9884} (\bibinfo {year} {2005})},\ \bibinfo {note} {pMID:
  16852194}\BibitemShut {NoStop}%
\bibitem [{\citenamefont {Booth}(1951)}]{Booth1951Dielectric}%
  \BibitemOpen
  \bibfield  {author} {\bibinfo {author} {\bibfnamefont {F.}~\bibnamefont
  {Booth}},\ }\bibfield  {title} {\enquote {\bibinfo {title} {The dielectric
  constant of water and the saturation effect},}\ }\href@noop {} {\bibfield
  {journal} {\bibinfo  {journal} {The Journal of Chemical Physics}\ }\textbf
  {\bibinfo {volume} {19}},\ \bibinfo {pages} {391--394} (\bibinfo {year}
  {1951})}\BibitemShut {NoStop}%
\bibitem [{\citenamefont {Piekara}\ and\ \citenamefont
  {Kielich}(1958)}]{Piekara1958}%
  \BibitemOpen
  \bibfield  {author} {\bibinfo {author} {\bibfnamefont {A.}~\bibnamefont
  {Piekara}}\ and\ \bibinfo {author} {\bibfnamefont {S.}~\bibnamefont
  {Kielich}},\ }\bibfield  {title} {\enquote {\bibinfo {title} {{Theory of
  Orientational Effects and Related Phenomena in Dielectric Liquids}},}\ }\href
  {\doibase 10.1063/1.1744712} {\bibfield  {journal} {\bibinfo  {journal} {The
  Journal of Chemical Physics}\ }\textbf {\bibinfo {volume} {29}},\ \bibinfo
  {pages} {1297--1305} (\bibinfo {year} {1958})}\BibitemShut {NoStop}%
\bibitem [{\citenamefont {Yeh}\ and\ \citenamefont
  {Berkowitz}(1999)}]{Yeh1999Dielectric}%
  \BibitemOpen
  \bibfield  {author} {\bibinfo {author} {\bibfnamefont {I.-C.}\ \bibnamefont
  {Yeh}}\ and\ \bibinfo {author} {\bibfnamefont {M.~L.}\ \bibnamefont
  {Berkowitz}},\ }\bibfield  {title} {\enquote {\bibinfo {title} {{Dielectric
  constant of water at high electric fields: Molecular dynamics study}},}\
  }\href {\doibase 10.1063/1.478698} {\bibfield  {journal} {\bibinfo  {journal}
  {The Journal of Chemical Physics}\ }\textbf {\bibinfo {volume} {110}},\
  \bibinfo {pages} {7935--7942} (\bibinfo {year} {1999})}\BibitemShut {NoStop}%
\bibitem [{\citenamefont {Kotodziej}, \citenamefont {Jones},\ and\
  \citenamefont {Davies}(1975)}]{Kotodziej1975High}%
  \BibitemOpen
  \bibfield  {author} {\bibinfo {author} {\bibfnamefont {H.~A.}\ \bibnamefont
  {Kotodziej}}, \bibinfo {author} {\bibfnamefont {G.~P.}\ \bibnamefont
  {Jones}}, \ and\ \bibinfo {author} {\bibfnamefont {M.}~\bibnamefont
  {Davies}},\ }\bibfield  {title} {\enquote {\bibinfo {title} {High field
  dielectric measurements in water},}\ }\href {\doibase 10.1039/F29757100269}
  {\bibfield  {journal} {\bibinfo  {journal} {J. Chem. Soc.{,} Faraday Trans.
  2}\ }\textbf {\bibinfo {volume} {71}},\ \bibinfo {pages} {269--274} (\bibinfo
  {year} {1975})}\BibitemShut {NoStop}%
\bibitem [{\citenamefont {Andrade}\ and\ \citenamefont
  {Dodd}(1946)}]{Andrade1946}%
  \BibitemOpen
  \bibfield  {author} {\bibinfo {author} {\bibfnamefont {E.~N. D.~C.}\
  \bibnamefont {Andrade}}\ and\ \bibinfo {author} {\bibfnamefont
  {C.}~\bibnamefont {Dodd}},\ }\bibfield  {title} {\enquote {\bibinfo {title}
  {The effect of an electric field on the viscosity of liquids},}\ }\href
  {\doibase 10.1098/rspa.1946.0079} {\bibfield  {journal} {\bibinfo  {journal}
  {Proceedings of the Royal Society of London. Series A. Mathematical and
  Physical Sciences}\ }\textbf {\bibinfo {volume} {187}},\ \bibinfo {pages}
  {296--337} (\bibinfo {year} {1946})}\BibitemShut {NoStop}%
\bibitem [{\citenamefont {Lyklema}\ and\ \citenamefont
  {Overbeek}(1961)}]{Lyklema1961}%
  \BibitemOpen
  \bibfield  {author} {\bibinfo {author} {\bibfnamefont {J.}~\bibnamefont
  {Lyklema}}\ and\ \bibinfo {author} {\bibfnamefont {J.}~\bibnamefont
  {Overbeek}},\ }\bibfield  {title} {\enquote {\bibinfo {title} {On the
  interpretation of electrokinetic potentials},}\ }\href {\doibase
  https://doi.org/10.1016/0095-8522(61)90029-0} {\bibfield  {journal} {\bibinfo
   {journal} {Journal of Colloid Science}\ }\textbf {\bibinfo {volume} {16}},\
  \bibinfo {pages} {501--512} (\bibinfo {year} {1961})}\BibitemShut {NoStop}%
\bibitem [{\citenamefont {Zong}\ \emph {et~al.}(2016)\citenamefont {Zong},
  \citenamefont {Hu}, \citenamefont {Duan},\ and\ \citenamefont
  {Sun}}]{Zong2016Viscosity}%
  \BibitemOpen
  \bibfield  {author} {\bibinfo {author} {\bibfnamefont {D.}~\bibnamefont
  {Zong}}, \bibinfo {author} {\bibfnamefont {H.}~\bibnamefont {Hu}}, \bibinfo
  {author} {\bibfnamefont {Y.}~\bibnamefont {Duan}}, \ and\ \bibinfo {author}
  {\bibfnamefont {Y.}~\bibnamefont {Sun}},\ }\bibfield  {title} {\enquote
  {\bibinfo {title} {Viscosity of water under electric field: Anisotropy
  induced by redistribution of hydrogen bonds},}\ }\href {\doibase
  10.1021/acs.jpcb.6b01686} {\bibfield  {journal} {\bibinfo  {journal} {J.
  Phys. Chem. B}\ }\textbf {\bibinfo {volume} {120}},\ \bibinfo {pages}
  {4818--4827} (\bibinfo {year} {2016})}\BibitemShut {NoStop}%
\bibitem [{\citenamefont {Jin}\ \emph {et~al.}(2022)\citenamefont {Jin},
  \citenamefont {Hwang}, \citenamefont {Chai}, \citenamefont {Kampf},\ and\
  \citenamefont {Klein}}]{Jin2022Direct}%
  \BibitemOpen
  \bibfield  {author} {\bibinfo {author} {\bibfnamefont {D.}~\bibnamefont
  {Jin}}, \bibinfo {author} {\bibfnamefont {Y.}~\bibnamefont {Hwang}}, \bibinfo
  {author} {\bibfnamefont {L.}~\bibnamefont {Chai}}, \bibinfo {author}
  {\bibfnamefont {N.}~\bibnamefont {Kampf}}, \ and\ \bibinfo {author}
  {\bibfnamefont {J.}~\bibnamefont {Klein}},\ }\bibfield  {title} {\enquote
  {\bibinfo {title} {Direct measurement of the viscoelectric effect in
  water},}\ }\href {\doibase 10.1073/pnas.2113690119} {\bibfield  {journal}
  {\bibinfo  {journal} {Proceedings of the National Academy of Sciences}\
  }\textbf {\bibinfo {volume} {119}},\ \bibinfo {pages} {e2113690119} (\bibinfo
  {year} {2022})}\BibitemShut {NoStop}%
\bibitem [{\citenamefont {Abrashkin}, \citenamefont {Andelman},\ and\
  \citenamefont {Orland}(2007)}]{Abrashkin2007Dipolar}%
  \BibitemOpen
  \bibfield  {author} {\bibinfo {author} {\bibfnamefont {A.}~\bibnamefont
  {Abrashkin}}, \bibinfo {author} {\bibfnamefont {D.}~\bibnamefont {Andelman}},
  \ and\ \bibinfo {author} {\bibfnamefont {H.}~\bibnamefont {Orland}},\
  }\bibfield  {title} {\enquote {\bibinfo {title} {Dipolar poisson-boltzmann
  equation: Ions and dipoles close to charge interfaces},}\ }\href {\doibase
  10.1103/PhysRevLett.99.077801} {\bibfield  {journal} {\bibinfo  {journal}
  {Phys. Rev. Lett.}\ }\textbf {\bibinfo {volume} {99}},\ \bibinfo {pages}
  {077801} (\bibinfo {year} {2007})}\BibitemShut {NoStop}%
\bibitem [{\citenamefont {Berendsen}, \citenamefont {Grigera},\ and\
  \citenamefont {Straatsma}(1987)}]{Berendsen1987Missing}%
  \BibitemOpen
  \bibfield  {author} {\bibinfo {author} {\bibfnamefont {H.~J.~C.}\
  \bibnamefont {Berendsen}}, \bibinfo {author} {\bibfnamefont {J.~R.}\
  \bibnamefont {Grigera}}, \ and\ \bibinfo {author} {\bibfnamefont {T.~P.}\
  \bibnamefont {Straatsma}},\ }\bibfield  {title} {\enquote {\bibinfo {title}
  {The missing term in effective pair potentials},}\ }\href {\doibase
  10.1021/j100308a038} {\bibfield  {journal} {\bibinfo  {journal} {J. Phys.
  Chem.}\ }\textbf {\bibinfo {volume} {91}},\ \bibinfo {pages} {6269--6271}
  (\bibinfo {year} {1987})}\BibitemShut {NoStop}%
\bibitem [{\citenamefont {Smith}\ and\ \citenamefont {{van
  Gunsteren}}(1993)}]{Smith1993Viscosity}%
  \BibitemOpen
  \bibfield  {author} {\bibinfo {author} {\bibfnamefont {P.~E.}\ \bibnamefont
  {Smith}}\ and\ \bibinfo {author} {\bibfnamefont {W.~F.}\ \bibnamefont {{van
  Gunsteren}}},\ }\bibfield  {title} {\enquote {\bibinfo {title} {The viscosity
  of spc and spc/e water at 277 and 300 k},}\ }\href {\doibase
  https://doi.org/10.1016/0009-2614(93)85720-9} {\bibfield  {journal} {\bibinfo
   {journal} {Chemical Physics Letters}\ }\textbf {\bibinfo {volume} {215}},\
  \bibinfo {pages} {315--318} (\bibinfo {year} {1993})}\BibitemShut {NoStop}%
\bibitem [{\citenamefont {Loche}\ \emph {et~al.}(2021)\citenamefont {Loche},
  \citenamefont {Steinbrunner}, \citenamefont {Friedowitz}, \citenamefont
  {Netz},\ and\ \citenamefont {Bonthuis}}]{Loche2021Transferable}%
  \BibitemOpen
  \bibfield  {author} {\bibinfo {author} {\bibfnamefont {P.}~\bibnamefont
  {Loche}}, \bibinfo {author} {\bibfnamefont {P.}~\bibnamefont {Steinbrunner}},
  \bibinfo {author} {\bibfnamefont {S.}~\bibnamefont {Friedowitz}}, \bibinfo
  {author} {\bibfnamefont {R.~R.}\ \bibnamefont {Netz}}, \ and\ \bibinfo
  {author} {\bibfnamefont {D.~J.}\ \bibnamefont {Bonthuis}},\ }\bibfield
  {title} {\enquote {\bibinfo {title} {Transferable ion force fields in water
  from a simultaneous optimization of ion solvation and ion-ion interaction},}\
  }\href {\doibase 10.1021/acs.jpcb.1c05303} {\bibfield  {journal} {\bibinfo
  {journal} {J. Phys. Chem. B}\ }\textbf {\bibinfo {volume} {125}},\ \bibinfo
  {pages} {8581--8587} (\bibinfo {year} {2021})}\BibitemShut {NoStop}%
\bibitem [{\citenamefont {Hess}\ \emph {et~al.}(2008)\citenamefont {Hess},
  \citenamefont {Kutzner}, \citenamefont {van~der Spoel},\ and\ \citenamefont
  {Lindahl}}]{Hess2008Gromacs4}%
  \BibitemOpen
  \bibfield  {author} {\bibinfo {author} {\bibfnamefont {B.}~\bibnamefont
  {Hess}}, \bibinfo {author} {\bibfnamefont {C.}~\bibnamefont {Kutzner}},
  \bibinfo {author} {\bibfnamefont {D.}~\bibnamefont {van~der Spoel}}, \ and\
  \bibinfo {author} {\bibfnamefont {E.}~\bibnamefont {Lindahl}},\ }\bibfield
  {title} {\enquote {\bibinfo {title} {Gromacs 4: Algorithms for highly
  efficient, load-balanced, and scalable molecular simulation},}\ }\href
  {\doibase 10.1021/ct700301q} {\bibfield  {journal} {\bibinfo  {journal} {J.
  Chem. Theory Comput.}\ }\textbf {\bibinfo {volume} {4}},\ \bibinfo {pages}
  {435--447} (\bibinfo {year} {2008})}\BibitemShut {NoStop}%
\bibitem [{Git()}]{GitHub}%
  \BibitemOpen
  \href {https://github.com/acmaggs/Wien} {\enquote {\bibinfo {title} {{See
  this GitHub repository for the simulation files and the Python codes used to
  interpret the simulations results.}}}\ }\BibitemShut {NoStop}%
\bibitem [{\citenamefont {Fulton}(2009)}]{Fulton2009Nonlinear}%
  \BibitemOpen
  \bibfield  {author} {\bibinfo {author} {\bibfnamefont {R.~L.}\ \bibnamefont
  {Fulton}},\ }\bibfield  {title} {\enquote {\bibinfo {title} {{The nonlinear
  dielectric behavior of water: Comparisons of various approaches to the
  nonlinear dielectric increment}},}\ }\href {\doibase 10.1063/1.3139211}
  {\bibfield  {journal} {\bibinfo  {journal} {The Journal of Chemical Physics}\
  }\textbf {\bibinfo {volume} {130}},\ \bibinfo {pages} {204503} (\bibinfo
  {year} {2009})}\BibitemShut {NoStop}%
\bibitem [{\citenamefont {Berthoumieux}, \citenamefont {Monet},\ and\
  \citenamefont {Blossey}(2021)}]{monetJCP2021}%
  \BibitemOpen
  \bibfield  {author} {\bibinfo {author} {\bibfnamefont {H.}~\bibnamefont
  {Berthoumieux}}, \bibinfo {author} {\bibfnamefont {G.}~\bibnamefont {Monet}},
  \ and\ \bibinfo {author} {\bibfnamefont {R.}~\bibnamefont {Blossey}},\
  }\bibfield  {title} {\enquote {\bibinfo {title} {{Dipolar Poisson models in a
  dual view}},}\ }\href
  {{https://pubs.aip.org/aip/jcp/article/155/2/024112/1064992}} {\bibfield
  {journal} {\bibinfo  {journal} {{J. Chem Phys.}}\ }\textbf {\bibinfo {volume}
  {155}},\ \bibinfo {pages} {024112} (\bibinfo {year} {2021})}\BibitemShut
  {NoStop}%
\bibitem [{\citenamefont {Landau}\ \emph {et~al.}(2013)\citenamefont {Landau},
  \citenamefont {Bell}, \citenamefont {Kearsley}, \citenamefont {Pitaevskii},
  \citenamefont {Lifshitz},\ and\ \citenamefont
  {Sykes}}]{landau2013electrodynamics}%
  \BibitemOpen
  \bibfield  {author} {\bibinfo {author} {\bibfnamefont {L.~D.}\ \bibnamefont
  {Landau}}, \bibinfo {author} {\bibfnamefont {J.~S.}\ \bibnamefont {Bell}},
  \bibinfo {author} {\bibfnamefont {M.}~\bibnamefont {Kearsley}}, \bibinfo
  {author} {\bibfnamefont {L.}~\bibnamefont {Pitaevskii}}, \bibinfo {author}
  {\bibfnamefont {E.}~\bibnamefont {Lifshitz}}, \ and\ \bibinfo {author}
  {\bibfnamefont {J.}~\bibnamefont {Sykes}},\ }\href@noop {} {\emph {\bibinfo
  {title} {Electrodynamics of continuous media}}},\ Vol.~\bibinfo {volume} {8}\
  (\bibinfo  {publisher} {elsevier},\ \bibinfo {year} {2013})\BibitemShut
  {NoStop}%
\bibitem [{\citenamefont {Hasted}, \citenamefont {Ritson},\ and\ \citenamefont
  {Collie}(1948)}]{hasted48}%
  \BibitemOpen
  \bibfield  {author} {\bibinfo {author} {\bibfnamefont {J.~B.}\ \bibnamefont
  {Hasted}}, \bibinfo {author} {\bibfnamefont {D.~M.}\ \bibnamefont {Ritson}},
  \ and\ \bibinfo {author} {\bibfnamefont {C.~H.}\ \bibnamefont {Collie}},\
  }\bibfield  {title} {\enquote {\bibinfo {title} {Dielectric properties of
  aqueous ionic solutions. parts \textrm{I} and \textrm{II}.}}\ }\href
  {{https://pubs.aip.org/aip/jcp/article-abstract/16/1/1/199390/Dielectric-Properties-of-Aqueous-Ionic-Solutions}}
  {\bibfield  {journal} {\bibinfo  {journal} {J. Chem. Phys.}\ }\textbf
  {\bibinfo {volume} {16}},\ \bibinfo {pages} {1--21} (\bibinfo {year}
  {1948})}\BibitemShut {NoStop}%
\bibitem [{\citenamefont {Levy}, \citenamefont {Andelman},\ and\ \citenamefont
  {Orland}(2012)}]{levy2012}%
  \BibitemOpen
  \bibfield  {author} {\bibinfo {author} {\bibfnamefont {A.}~\bibnamefont
  {Levy}}, \bibinfo {author} {\bibfnamefont {D.}~\bibnamefont {Andelman}}, \
  and\ \bibinfo {author} {\bibfnamefont {H.}~\bibnamefont {Orland}},\
  }\bibfield  {title} {\enquote {\bibinfo {title} {Dielectric constant of ionic
  solutions: A field-theory approach},}\ }\href
  {{https://journals.aps.org/prl/abstract/10.1103/PhysRevLett.108.227801}}
  {\bibfield  {journal} {\bibinfo  {journal} {Phys. Rev. Lett.}\ }\textbf
  {\bibinfo {volume} {108}},\ \bibinfo {pages} {227801} (\bibinfo {year}
  {2012})}\BibitemShut {NoStop}%
\bibitem [{\citenamefont {Seal}, \citenamefont {Doblhoff-Dier},\ and\
  \citenamefont {Meyer}(2019)}]{seal2019}%
  \BibitemOpen
  \bibfield  {author} {\bibinfo {author} {\bibfnamefont {S.}~\bibnamefont
  {Seal}}, \bibinfo {author} {\bibfnamefont {K.}~\bibnamefont {Doblhoff-Dier}},
  \ and\ \bibinfo {author} {\bibfnamefont {J.}~\bibnamefont {Meyer}},\
  }\bibfield  {title} {\enquote {\bibinfo {title} {Dielectric decrement for
  aqueous nacl solutions: Effect of ionic charge scaling in nonpolarizable
  water force fields},}\ }\href
  {{https://journals.aps.org/prl/abstract/10.1103/PhysRevLett.108.227801}}
  {\bibfield  {journal} {\bibinfo  {journal} {J. Phys. Chem. B}\ }\textbf
  {\bibinfo {volume} {123}},\ \bibinfo {pages} {9912--9921} (\bibinfo {year}
  {2019})}\BibitemShut {NoStop}%
\bibitem [{\citenamefont {Bonneau}, \citenamefont {Démery},\ and\
  \citenamefont {Raphaël}(2023)}]{Bonneau2023Temporal}%
  \BibitemOpen
  \bibfield  {author} {\bibinfo {author} {\bibfnamefont {H.}~\bibnamefont
  {Bonneau}}, \bibinfo {author} {\bibfnamefont {V.}~\bibnamefont {Démery}}, \
  and\ \bibinfo {author} {\bibfnamefont {E.}~\bibnamefont {Raphaël}},\
  }\bibfield  {title} {\enquote {\bibinfo {title} {Temporal response of the
  conductivity of electrolytes},}\ }\href {\doibase 10.1088/1742-5468/acdced}
  {\bibfield  {journal} {\bibinfo  {journal} {Journal of Statistical Mechanics:
  Theory and Experiment}\ }\textbf {\bibinfo {volume} {2023}},\ \bibinfo
  {pages} {073205} (\bibinfo {year} {2023})}\BibitemShut {NoStop}%
\bibitem [{\citenamefont {Bopp}, \citenamefont {Kornyshev},\ and\ \citenamefont
  {Sutmann}(1996)}]{Bopp1996}%
  \BibitemOpen
  \bibfield  {author} {\bibinfo {author} {\bibfnamefont {P.~A.}\ \bibnamefont
  {Bopp}}, \bibinfo {author} {\bibfnamefont {A.~A.}\ \bibnamefont {Kornyshev}},
  \ and\ \bibinfo {author} {\bibfnamefont {G.}~\bibnamefont {Sutmann}},\
  }\bibfield  {title} {\enquote {\bibinfo {title} {{Static Nonlocal Dielectric
  Function of Liquid Water}},}\ }\href {\doibase 10.1103/PhysRevLett.76.1280}
  {\bibfield  {journal} {\bibinfo  {journal} {{Phys. Rev. Lett.}}\ }\textbf
  {\bibinfo {volume} {76}},\ \bibinfo {pages} {1280--1283} (\bibinfo {year}
  {1996})}\BibitemShut {NoStop}%
\bibitem [{\citenamefont {Berthoumieux}(2018)}]{Berthoumieux2018Gaussian}%
  \BibitemOpen
  \bibfield  {author} {\bibinfo {author} {\bibfnamefont {H.}~\bibnamefont
  {Berthoumieux}},\ }\bibfield  {title} {\enquote {\bibinfo {title} {Gaussian
  field model for polar fluids as a function of density and polarization:
  Toward a model for water},}\ }\href {\doibase 10.1063/1.5012828} {\bibfield
  {journal} {\bibinfo  {journal} {The Journal of Chemical Physics}\ }\textbf
  {\bibinfo {volume} {148}},\ \bibinfo {pages} {104504} (\bibinfo {year}
  {2018})}\BibitemShut {NoStop}%
\bibitem [{\citenamefont {Bopp}, \citenamefont {Kornyshev},\ and\ \citenamefont
  {Sutmann}(1998)}]{Bopp1998}%
  \BibitemOpen
  \bibfield  {author} {\bibinfo {author} {\bibfnamefont {P.~A.}\ \bibnamefont
  {Bopp}}, \bibinfo {author} {\bibfnamefont {A.~A.}\ \bibnamefont {Kornyshev}},
  \ and\ \bibinfo {author} {\bibfnamefont {G.}~\bibnamefont {Sutmann}},\
  }\bibfield  {title} {\enquote {\bibinfo {title} {{Frequency and wave-vector
  dependent dielectric function of water: Collective modes and relaxation
  spectra}},}\ }\href {\doibase 10.1063/1.476884} {\bibfield  {journal}
  {\bibinfo  {journal} {The Journal of Chemical Physics}\ }\textbf {\bibinfo
  {volume} {109}},\ \bibinfo {pages} {1939--1958} (\bibinfo {year}
  {1998})}\BibitemShut {NoStop}%
\bibitem [{\citenamefont {Illien}, \citenamefont {Carof},\ and\ \citenamefont
  {Rotenberg}(2024)}]{Illien2024Stochastic}%
  \BibitemOpen
  \bibfield  {author} {\bibinfo {author} {\bibfnamefont {P.}~\bibnamefont
  {Illien}}, \bibinfo {author} {\bibfnamefont {A.}~\bibnamefont {Carof}}, \
  and\ \bibinfo {author} {\bibfnamefont {B.}~\bibnamefont {Rotenberg}},\ }\href
  {https://arxiv.org/abs/2407.17232} {\enquote {\bibinfo {title} {Stochastic
  density functional theory for ions in a polar solvent},}\ } (\bibinfo {year}
  {2024}),\ \Eprint {http://arxiv.org/abs/2407.17232} {arXiv:2407.17232
  [cond-mat.soft]} \BibitemShut {NoStop}%
\bibitem [{\citenamefont {Jalali}\ \emph {et~al.}(2021)\citenamefont {Jalali},
  \citenamefont {Lofte}, \citenamefont {Boya},\ and\ \citenamefont
  {Neek-Amal}}]{Jalali2021}%
  \BibitemOpen
  \bibfield  {author} {\bibinfo {author} {\bibfnamefont {H.}~\bibnamefont
  {Jalali}}, \bibinfo {author} {\bibfnamefont {E.}~\bibnamefont {Lofte}},
  \bibinfo {author} {\bibfnamefont {R.}~\bibnamefont {Boya}}, \ and\ \bibinfo
  {author} {\bibfnamefont {M.}~\bibnamefont {Neek-Amal}},\ }\bibfield  {title}
  {\enquote {\bibinfo {title} {{Abnormal Dielectric Constant of Nanoconﬁned
  Water between Graphene Layers in the Presence of Salt}},}\ }\href
  {{https://pubs.acs.org/doi/10.1021/acs.jpcb.0c09156}} {\bibfield  {journal}
  {\bibinfo  {journal} {{J. Phys. Chem. B}}\ }\textbf {\bibinfo {volume}
  {125}},\ \bibinfo {pages} {1604--1610} (\bibinfo {year} {2021})}\BibitemShut
  {NoStop}%
\bibitem [{\citenamefont {Loche}\ \emph {et~al.}(2019)\citenamefont {Loche},
  \citenamefont {Ayaz}, \citenamefont {Schlaich}, \citenamefont {Uematsu},\
  and\ \citenamefont {Netz}}]{Loche2019Giant}%
  \BibitemOpen
  \bibfield  {author} {\bibinfo {author} {\bibfnamefont {P.}~\bibnamefont
  {Loche}}, \bibinfo {author} {\bibfnamefont {C.}~\bibnamefont {Ayaz}},
  \bibinfo {author} {\bibfnamefont {A.}~\bibnamefont {Schlaich}}, \bibinfo
  {author} {\bibfnamefont {Y.}~\bibnamefont {Uematsu}}, \ and\ \bibinfo
  {author} {\bibfnamefont {R.~R.}\ \bibnamefont {Netz}},\ }\bibfield  {title}
  {\enquote {\bibinfo {title} {{Giant Axial Dielectric Response in Water-Filled
  Nanotubes and Effective Electrostatic Ion–Ion Interactions from a Tensorial
  Dielectric Model}},}\ }\href {\doibase 10.1021/acs.jpcb.9b09269} {\bibfield
  {journal} {\bibinfo  {journal} {{J. Phys. Chem. B}}\ }\textbf {\bibinfo
  {volume} {123}},\ \bibinfo {pages} {10850--10857} (\bibinfo {year}
  {2019})}\BibitemShut {NoStop}%
\bibitem [{\citenamefont {Wang}\ \emph {et~al.}(2024)\citenamefont {Wang},
  \citenamefont {Souilamas}, \citenamefont {Esfandiar}, \citenamefont
  {Fabregas}, \citenamefont {Benaglia}, \citenamefont {Nevison-Andrews},
  \citenamefont {Yang}, \citenamefont {Normansell}, \citenamefont {Ares},
  \citenamefont {Ferrari}, \citenamefont {Principi}, \citenamefont {Geim},\
  and\ \citenamefont {Fumagalli}}]{Wang2024}%
  \BibitemOpen
  \bibfield  {author} {\bibinfo {author} {\bibfnamefont {R.}~\bibnamefont
  {Wang}}, \bibinfo {author} {\bibfnamefont {M.}~\bibnamefont {Souilamas}},
  \bibinfo {author} {\bibfnamefont {A.}~\bibnamefont {Esfandiar}}, \bibinfo
  {author} {\bibfnamefont {R.}~\bibnamefont {Fabregas}}, \bibinfo {author}
  {\bibfnamefont {S.}~\bibnamefont {Benaglia}}, \bibinfo {author}
  {\bibfnamefont {H.}~\bibnamefont {Nevison-Andrews}}, \bibinfo {author}
  {\bibfnamefont {Q.}~\bibnamefont {Yang}}, \bibinfo {author} {\bibfnamefont
  {J.}~\bibnamefont {Normansell}}, \bibinfo {author} {\bibfnamefont
  {P.}~\bibnamefont {Ares}}, \bibinfo {author} {\bibfnamefont {G.}~\bibnamefont
  {Ferrari}}, \bibinfo {author} {\bibfnamefont {A.}~\bibnamefont {Principi}},
  \bibinfo {author} {\bibfnamefont {A.~K.}\ \bibnamefont {Geim}}, \ and\
  \bibinfo {author} {\bibfnamefont {L.}~\bibnamefont {Fumagalli}},\ }\href
  {https://arxiv.org/abs/2407.21538} {\enquote {\bibinfo {title} {In-plane
  dielectric constant and conductivity of confined water},}\ } (\bibinfo {year}
  {2024}),\ \Eprint {http://arxiv.org/abs/2407.21538} {arXiv:2407.21538
  [cond-mat.mes-hall]} \BibitemShut {NoStop}%
\bibitem [{\citenamefont {Fumagalli}\ \emph {et~al.}(2018)\citenamefont
  {Fumagalli}, \citenamefont {Esfandiar}, \citenamefont {Fabregas},
  \citenamefont {Hu}, \citenamefont {Ares}, \citenamefont {Janardanan},
  \citenamefont {Yang}, \citenamefont {Rhada}, \citenamefont {Tanigushi},
  \citenamefont {Watanabe}, \citenamefont {Gomila}, \citenamefont {Novoselov},\
  and\ \citenamefont {Geim}}]{fumagalli2018}%
  \BibitemOpen
  \bibfield  {author} {\bibinfo {author} {\bibfnamefont {L.}~\bibnamefont
  {Fumagalli}}, \bibinfo {author} {\bibfnamefont {A.}~\bibnamefont
  {Esfandiar}}, \bibinfo {author} {\bibfnamefont {R.}~\bibnamefont {Fabregas}},
  \bibinfo {author} {\bibfnamefont {S.}~\bibnamefont {Hu}}, \bibinfo {author}
  {\bibfnamefont {P.}~\bibnamefont {Ares}}, \bibinfo {author} {\bibfnamefont
  {A.}~\bibnamefont {Janardanan}}, \bibinfo {author} {\bibfnamefont
  {Q.}~\bibnamefont {Yang}}, \bibinfo {author} {\bibfnamefont {B.}~\bibnamefont
  {Rhada}}, \bibinfo {author} {\bibfnamefont {T.}~\bibnamefont {Tanigushi}},
  \bibinfo {author} {\bibfnamefont {K.}~\bibnamefont {Watanabe}}, \bibinfo
  {author} {\bibfnamefont {G.}~\bibnamefont {Gomila}}, \bibinfo {author}
  {\bibfnamefont {K.~S.}\ \bibnamefont {Novoselov}}, \ and\ \bibinfo {author}
  {\bibfnamefont {A.~K.}\ \bibnamefont {Geim}},\ }\bibfield  {title} {\enquote
  {\bibinfo {title} {{Anomalous low dielectric constant of confined water}},}\
  }\href {{https://www.science.org/doi/full/10.1126/science.aat4191}}
  {\bibfield  {journal} {\bibinfo  {journal} {{Science}}\ }\textbf {\bibinfo
  {volume} {360}},\ \bibinfo {pages} {1339--1342} (\bibinfo {year}
  {2018})}\BibitemShut {NoStop}%
\bibitem [{\citenamefont {Neumann}(1983)}]{neuman83}%
  \BibitemOpen
  \bibfield  {author} {\bibinfo {author} {\bibfnamefont {M.}~\bibnamefont
  {Neumann}},\ }\bibfield  {title} {\enquote {\bibinfo {title} {Dipole moment
  fluctuation formulas in computer simulations of polar systems},}\ }\href
  {\doibase 10.1080/00268978300102721} {\bibfield  {journal} {\bibinfo
  {journal} {Mol. Phys.}\ }\textbf {\bibinfo {volume} {50}},\ \bibinfo {pages}
  {841--858} (\bibinfo {year} {1983})}\BibitemShut {NoStop}%
\bibitem [{\citenamefont {Flyvbjerg}\ and\ \citenamefont
  {Petersen}(1989)}]{Flyvbjerg1989Error}%
  \BibitemOpen
  \bibfield  {author} {\bibinfo {author} {\bibfnamefont {H.}~\bibnamefont
  {Flyvbjerg}}\ and\ \bibinfo {author} {\bibfnamefont {H.~G.}\ \bibnamefont
  {Petersen}},\ }\bibfield  {title} {\enquote {\bibinfo {title} {{Error
  estimates on averages of correlated data}},}\ }\href {\doibase
  10.1063/1.457480} {\bibfield  {journal} {\bibinfo  {journal} {The Journal of
  Chemical Physics}\ }\textbf {\bibinfo {volume} {91}},\ \bibinfo {pages}
  {461--466} (\bibinfo {year} {1989})}\BibitemShut {NoStop}%
\end{thebibliography}
%
\end{document}